\documentclass{myaa}  
\usepackage{natbib}
\usepackage{graphicx}
\usepackage{txfonts}
\begin{document}
\def\teff{$T\rm_{eff }$}
\def\kms{$\mathrm {km s}^{-1}$}
\def\gtsima
{\hbox{\raise0.5ex\hbox{$>\lower1.06ex\hbox{$\kern-1.07em{\sim}$}$}}}
\def\ltsima
{\hbox{\raise0.5ex\hbox{$<\lower1.06ex\hbox{$\kern-1.07em{\sim}$}$}}}

\title{High-resolution spectroscopy of \\ RGB stars in the Sagittarius
Streams \thanks{Based on observations taken at ESO VLT Kueyen telescope (Cerro Paranal,
Chile, program: 075.B-0127(A)) and 3.6m telescope (La Silla, Chile). Also based
on spectroscopic observations  taken at the Telescopio Nazionale Galileo,
operated by the Fundaci\'on G. Galilei of INAF at  the Spanish Observatorio del
Roque de los Muchachos of the IAC (La Palma, Spain).} 
}

   \subtitle{I. Radial velocities and chemical abundances
}

\author{
L. \,Monaco \inst{1},
M. \,Bellazzini \inst{2},
P. \,Bonifacio \inst{3,4,5},
A. \,Buzzoni \inst{2},\\
F.R. \,Ferraro \inst{6},
G. \,Marconi \inst{1},
L. \,Sbordone \inst{5,7},
\and S. \,Zaggia \inst{8}
          }

  \offprints{L. Monaco}

\institute{
European Southern Observatory, Casilla 19001, Santiago, Chile
\and
Istituto Nazionale di Astrofisica --
Osservatorio Astronomico di Bologna, Italy
I-40127 Bologna, Italy
\and
CIFIST Marie Curie Excellence Team
         \and
Observatoire de Paris, GEPI , 5, place Jules Janssen, 92195 Meudon, France
	 \and
Istituto Nazionale di Astrofisica --
Osservatorio Astronomico di Trieste, Via Tiepolo 11,
I-34131 Trieste, Italy
\and
Universit\`a di Bologna --
Dipartimento di Astronomia,
I-40127 Bologna, Italy
\and
Istituto Nazionale di Astrofisica --
Osservatorio Astronomico di Roma, 
Via Frascati~33, 00040 Monteporzio Catone, Roma, Italy
\and
Istituto Nazionale di Astrofisica -- 
Osservatorio Astronomico di Padova, 
Vicolo dell'Osservatorio 5, 35122 Padova, Italy
}

\authorrunning{Monaco et al.}
\mail{lmonaco@eso.org}

\titlerunning{RGB stars in the Sgr dSph Streams}

\date{Received  / Accepted }


\abstract
{The Sagittarius (Sgr) dwarf spheroidal galaxy is currently being disrupted 
under the strain of the Milky Way. A reliable reconstruction of Sgr star
formation history can only be obtained by combining core and stream 
information. }
{
We
present radial velocities for 67 stars belonging to the Sgr Stream. For 12
stars in the sample we also present iron (Fe) and $\alpha$-element (Mg, Ca)
abundances.
}
{
Spectra were secured using different high resolution facilities: UVES@VLT,
HARPS@3.6m, and SARG@TNG. Radial velocities are obtained through cross 
correlation with a template spectra.  Concerning chemical analysis, for the
various elements, selected line equivalent widths were measured and 
abundances computed using the  WIDTH code and
ATLAS model atmospheres.
}
{
The velocity dispersion of the trailing tail is found to be
$\sigma$=8.3$\pm$0.9~km~s$^{-1}$, i.e., significantly lower than in the core of
the Sgr galaxy and marginally lower than previous estimates in the same portion
of the  stream. Stream stars follow the same trend as Sgr main body stars in the
[$\alpha$/Fe] vs [Fe/H] plane. However, stars  are, on average, more metal poor
in the stream than in the main body. This effect is slightly stronger in stars
belonging to more ancient wraps of the stream, according to currently accepted
models of Sgr disruption. 
}
{}


\maketitle

\begin{table*}
\caption{Basic parameters of the program stars.  
Measured radial
velocities and spectra signal-to-noise ratios are also reported. 
Beside the 2MASS name we give our own identifier to the star, 
which is used consequently in the other tables and throughout the paper.
}
\label{coord}
\begin{center}
{
\tiny
\begin{tabular}{lrccrrrrrrr|c|rr}
\hline
\hline
\multicolumn{14}{c}{SARG sample}\\
\hline
\hline
\\
2MASS ID & Star & \multicolumn{2}{c}{$\alpha\;\; (J2000.0)\;\;\; \delta\;\; (J2000.0)\;\;\;$}&
l & b & $\Lambda_\odot$ &
$K_0$ & $(J-K)_0$ & E(B-V) & $d_\odot(kpc)$ & S/N & v$_{helio}$(km/s) & v$_{gsr}$(km/s)
\\
&&&&&&&&&&&(@650nm)&&\\
\hline
\\
\object{2MASS~J00170214+0104165} & \object{1}   &  00 \, 17 \, 02.1   & +01 \, 04 \, 16 & 105.2  &  -60.6  & 82.18  & 10.88 & 0.95 & 0.03 &  15.07& 19& -227.91$\pm$0.48 &-125.29 \\
\object{2MASS~J01225543+0451341} & \object{19}  &  01 \, 22 \, 55.4   & +04 \, 51 \, 34 & 137.5  &  -57.1  & 98.69  & 10.89 & 0.97 & 0.03 &  16.39& 29& -207.07$\pm$0.45 &-131.44 \\
\object{2MASS~J01295300+0055392} & \object{23}  &  01 \, 29 \, 53.0   & +00 \, 55 \, 39 & 142.7  &  -60.5  & 98.08  & 11.21 & 0.94 & 0.02 &  16.42& 29& -163.52$\pm$0.25 &-103.86 \\
\object{2MASS~J01305453+0351575} & \object{25}  &  01 \, 30 \, 54.5   & +03 \, 51 \, 57 & 141.5  &  -57.6  & 99.86  & 11.19 & 0.96 & 0.02 &  18.22& 27& -188.82$\pm$0.30 &-121.11 \\
\object{2MASS~J01492111+0231536} & \object{39}  &  01 \, 49 \, 21.1   & +02 \, 31 \, 53 & 150.4  &  -57.2  & 103.1  & 10.77 & 0.99 & 0.02 &  16.43& 32& -188.31$\pm$0.51 &-136.39 \\
\object{2MASS~J01520195+0020525} & \object{42}  &  01 \, 52 \, 01.9   & +00 \, 20 \, 52 & 153.3  &  -58.9  & 102.5  & 10.73 & 0.96 & 0.03 &  14.56& 33& -160.10$\pm$0.33 &-116.45 \\
\object{2MASS~J02224517+0039362} & \object{77}  &  02 \, 22 \, 45.1   & +00 \, 39 \, 36 & 165.0  &  -54.6  & 109.2  & 10.64 & 0.97 & 0.03 &  14.20& 37&  -45.94$\pm$0.34 &-21.79  \\
			  \\																  
\object{2MASS~J08491184+4819380} & \object{232} &  08 \, 49 \, 11.8   & +48 \, 19 \, 38 & 171.2  &   39.2  & 202.9  & 11.10 & 0.97 & 0.03 &  17.69& 18&  -27.77$\pm$0.20 & -2.85  \\
\object{2MASS~J09193349+4350034} & \object{242} &  09 \, 19 \, 33.5   & +43 \, 50 \, 03 & 176.9  &   44.6  & 208.9  & 10.31 & 0.95 & 0.02 &  11.48& 28&	-8.24$\pm$0.19 & -0.90  \\
\object{2MASS~J10055844+3132597} & \object{260} &  10 \, 05 \, 58.4   & +31 \, 33 \, 00 & 195.8  &   53.9  & 220.5  & 10.48 & 0.97 & 0.02 &  13.53&  8&  107.71$\pm$0.33 & 71.12  \\
\object{2MASS~J10413374+4033099} & \object{278} &  10 \, 41 \, 33.8   & +40 \, 33 \, 10 & 177.9  &   60.1  & 224.8  & 11.35 & 0.93 & 0.02 &  17.15& 17&	 30.69$\pm$0.37 & 36.58  \\
\object{2MASS~J10472976+2339508} & \object{281} &  10 \, 47 \, 29.8   & +23 \, 39 \, 51 & 213.0  &   61.9  & 231.6  & 10.13 & 0.94 & 0.03 &  10.07& 25&	 56.53$\pm$0.41 & -0.49  \\
\object{2MASS~J10504744+2748542} & \object{283} &  10 \, 50 \, 47.5   & +27 \, 48 \, 54 & 204.4  &   63.3  & 230.9  & 10.62 & 0.95 & 0.02 &  13.09& 15&  -25.32$\pm$0.48 & -65.83 \\
\\
\hline
\hline
\multicolumn{14}{c}{HARPS sample}\\
\hline
\hline
\\
\object{2MASS~J13091343+1215502}& \object{465}    & 13 \, 09 \, 13.4& +12 \, 15 \, 50& 319.5 & 74.6& 266.8& 10.35 & 0.99& 0.02 &13.51 & 2 & -30.57$\pm$0.011&  -62.14 \\
\object{2MASS~J14340509+0936258}& \object{677}    & 14 \, 34 \, 05.1& +09 \, 36 \, 26&   1.9 & 60.1& 286.0&  9.60 & 0.99& 0.03 & 9.64 & 2 & -73.55$\pm$0.011&  -59.13 \\
\object{2MASS~J09364419+0704484}& \object{452892} & 09 \, 36 \, 44.2& +07 \, 04 \, 48& 227.2 & 39.7& 219.5&  9.94 & 0.94& 0.04 &10.18 &12 & 118.53$\pm$0.008&  -12.52 \\
\object{2MASS~J09453656+0937568}& \object{459992} & 09 \, 45 \, 36.6& +09 \, 37 \, 57& 225.6 & 42.9& 221.2&  9.91 & 0.99& 0.02 &11.47 &12 & 336.10$\pm$0.008&  214.81 \\
\object{2MASS~J09564707+0645292}& \object{452060} & 09 \, 56 \, 47.1& +06 \, 45 \, 29& 230.9 & 43.9& 224.8&  9.82 & 0.95& 0.02 & 9.62 &12 & 100.57$\pm$0.008&  -28.53 \\
\object{2MASS~J11345053-0700511}& \object{421173} & 11 \, 34 \, 50.5& -07 \, 00 \, 51& 271.7 & 51.1& 255.3&  9.81 & 0.97& 0.04 &10.84 &11 &  69.26$\pm$0.008&  -70.60 \\
\object{2MASS~J11554802-0339068}& \object{427535} & 11 \, 55 \, 48.0& -03 \, 39 \, 07& 277.2 & 56.4& 258.6& 10.01 & 0.99& 0.02 &12.42 & 7 & -57.53$\pm$0.009&  -178.38\\
\object{2MASS~J14112205-0610129}& \object{423286} & 14 \, 11 \, 22.1& -06 \, 10 \, 13& 336.0 & 51.5& 289.7&  9.51 & 1.05& 0.03 &12.55 &12 &  41.10$\pm$0.008&  -7.04  \\
\\
\hline
\hline
\multicolumn{14}{c}{UVES sample}\\
\hline
\hline
\\			  									       
\object{2MASS~J00035283-1940468} & \object{1005} & 00 \, 03 \, 52.8 & -19 \, 40 \, 47  &   64.8  & -76.8  &  69.5  &  10.97 & 1.03 & 0.02  &  19.1  &31 &  -56.92$\pm$ 0.13 & -14.93  \\
\object{2MASS~J00131891-2301528} & \object{1006} & 00 \, 13 \, 18.9 & -23 \, 01 \, 53  &   56.2  & -80.4  &  70.1  &  10.59 & 1.00 & 0.02  &  14.5  &42 &  -41.90$\pm$ 0.05 & -15.82  \\
\object{2MASS~J00135624-1721554} & \object{1007} & 00 \, 13 \, 56.2 & -17 \, 21 \, 55  &   79.4  & -76.9  &  72.7  &  11.52 & 0.99 & 0.03  &  21.4  &53 &  -90.51$\pm$ 0.16 & -45.26  \\
\object{2MASS~J00223563-0512079} & \object{1008} & 00 \, 22 \, 35.6 & -05 \, 12 \, 08  &  104.3  & -67.0  &  80.0  &   9.81 & 1.08 & 0.03  &  13.9  &38 & -116.25$\pm$ 0.31 & -35.72  \\
\object{2MASS~J00264668-1526427} & \object{1009} & 00 \, 26 \, 46.7 & -15 \, 26 \, 43  &   95.6  & -77.0  &  76.3  &  10.82 & 0.95 & 0.03  &  13.2  &49 &  -70.01$\pm$ 0.07 & -25.09  \\
\object{2MASS~J00321654-1851113} & \object{1010} & 00 \, 32 \, 16.5 & -18 \, 51 \, 11  &   93.9  & -80.6  &  75.9  &  11.57 & 1.02 & 0.02  &  24.5  &29 &  -71.25$\pm$ 0.10 & -40.45  \\
\object{2MASS~J00390964-1322429} & \object{1011} & 00 \, 39 \, 09.6 & -13 \, 22 \, 43  &  110.5  & -76.0  &  79.9  &  10.50 & 1.12 & 0.02  &  21.9  &18 &  -95.24$\pm$ 0.22 & -50.22  \\
\object{2MASS~J00464414-0659259} & \object{1013} & 00 \, 46 \, 44.1 & -06 \, 59 \, 26  &  119.5  & -69.8  &  84.5  &  11.84 & 0.95 & 0.06  &  21.0  &49 & -122.67$\pm$ 0.51 & -61.05  \\
\object{2MASS~J00480460-1131551} & \object{1014} & 00 \, 48 \, 04.6 & -11 \, 31 \, 55  &  119.9  & -74.4  &  82.7  &  10.61 & 1.05 & 0.03  &  17.3  &47 & -100.88$\pm$ 0.46 & -54.74  \\
\object{2MASS~J00522982-1518360} & \object{1015} & 00 \, 52 \, 29.8 & -15 \, 18 \, 36  &  124.2  & -78.2  &  81.9  &  11.16 & 1.08 & 0.02  &  25.3  &21 &  -87.05$\pm$ 0.67 & -55.70  \\
\object{2MASS~J00532013-0529477} & \object{1016} & 00 \, 53 \, 20.1 & -05 \, 29 \, 48  &  124.2  & -68.4  &  86.7  &  10.46 & 1.01 & 0.04  &  14.0  &55 & -148.81$\pm$ 0.45 & -86.55  \\
\object{2MASS~J00542073-0449174} & \object{1017} & 00 \, 54 \, 20.7 & -04 \, 49 \, 17  &  124.8  & -67.7  &  87.3  &  10.51 & 1.01 & 0.05  &  14.3  &47 & -139.50$\pm$ 0.42 & -75.63  \\
\object{2MASS~J00563325-2154386} & \object{1018} & 00 \, 56 \, 33.3 & -21 \, 54 \, 39  &  135.8  & -84.7  &  79.5  &  10.74 & 1.06 & 0.02  &  19.2  &29 &  -56.01$\pm$ 0.09 & -48.64  \\
\object{2MASS~J01011934-1536343} & \object{1019} & 01 \, 01 \, 19.3 & -15 \, 36 \, 34  &  134.7  & -78.3  &  83.6  &  11.20 & 1.01 & 0.02  &  19.6  &52 &  -86.01$\pm$ 0.55 & -60.71  \\
\object{2MASS~J01015376-1015085} & \object{1020} & 01 \, 01 \, 53.8 & -10 \, 15 \, 08  &  131.7  & -72.9  &  86.3  &  11.31 & 1.01 & 0.03  &  21.0  &63 & -119.20$\pm$ 0.53 & -76.72  \\
\object{2MASS~J01091912-1508157} & \object{1022} & 01 \, 09 \, 19.1 & -15 \, 08 \, 16  &  143.0  & -77.3  &  85.5  &   9.27 & 1.14 & 0.03  &  13.6  &18 & -102.48$\pm$ 0.43 & -80.19  \\
\object{2MASS~J01212317-1036096} & \object{1025} & 01 \, 21 \, 23.2 & -10 \, 36 \, 10  &  147.5  & -72.0  &  90.3  &  10.06 & 1.08 & 0.03  &  15.5  &38 & -139.46$\pm$ 0.16 & -109.94 \\
\object{2MASS~J01282756-0505173} & \object{1028} & 01 \, 28 \, 27.6 & -05 \, 05 \, 17  &  146.4  & -66.3  &  94.6  &  11.33 & 0.96 & 0.04  &  17.1  &71 & -123.43$\pm$ 0.52 & -81.25  \\
\object{2MASS~J01512105-0727451} & \object{1034} & 01 \, 51 \, 21.1 & -07 \, 27 \, 45  &  161.5  & -65.7  &  98.3  &  11.69 & 1.01 & 0.02  &  24.4  &52 & -129.68$\pm$ 0.37 & -109.28 \\
\object{2MASS~J01574156-1709471} & \object{1035} & 01 \, 57 \, 41.6 & -17 \, 09 \, 47  &  183.3  & -71.7  &  94.6  &  10.65 & 1.05 & 0.02  &  17.5  &28 &  -91.39$\pm$ 0.11 & -105.05 \\
\object{2MASS~J01593606-0801131} & \object{1036} & 01 \, 59 \, 36.1 & -08 \, 01 \, 13  &  166.3  & -65.0  &  99.8  &  10.66 & 1.05 & 0.02  &  18.0  &32 & -109.87$\pm$ 0.27 & -96.69  \\
\object{2MASS~J21153360-3530060} & \object{1065} & 21 \, 15 \, 33.6 & -35 \, 30 \, 06  &    8.5  & -43.7  &  29.5  &  11.73 & 1.03 & 0.07  &  26.7  &47 &   90.93$\pm$ 0.22 &  117.32 \\
\object{2MASS~J21174714-2432331} & \object{1066} & 21 \, 17 \, 47.1 & -24 \, 32 \, 33  &   23.4  & -42.2  &  31.5  &  11.92 & 1.01 & 0.05  &  27.4  &44 &   30.30$\pm$ 0.03 &  99.97  \\
\object{2MASS~J21250780-2747090} & \object{1067} & 21 \, 25 \, 07.8 & -27 \, 47 \, 09  &   19.6  & -44.6  &  32.6  &  10.65 & 1.00 & 0.10  &  14.5  &38 &   52.99$\pm$ 0.21 &  109.52 \\
\object{2MASS~J21314538-3513510} & \object{1068} & 21 \, 31 \, 45.4 & -35 \, 13 \, 51  &    9.3  & -47.0  &  32.8  &  11.65 & 1.01 & 0.07  &  24.0  &48 &   81.00$\pm$ 0.14 &  107.51 \\
\object{2MASS~J21581991-3406074} & \object{1071} & 21 \, 58 \, 19.9 & -34 \, 06 \, 07  &   11.4  & -52.4  &  38.4  &  10.74 & 1.07 & 0.02  &  20.0  &61 &   49.80$\pm$ 0.05 &  77.62  \\
\object{2MASS~J22083965-2812124} & \object{1072} & 22 \, 08 \, 39.7 & -28 \, 12 \, 12  &   21.5  & -54.1  &  42.0  &  10.92 & 1.04 & 0.02  &  19.1  &35 &   34.80$\pm$ 0.28 &  83.89  \\
\object{2MASS~J22142679-2306184} & \object{1073} & 22 \, 14 \, 26.8 & -23 \, 06 \, 18  &   30.5  & -54.4  &  44.5  &  10.24 & 1.10 & 0.03  &  17.8  &23 &  -11.35$\pm$ 0.40 &  56.01  \\
\object{2MASS~J22264953-3918313} & \object{1075} & 22 \, 26 \, 49.5 & -39 \, 18 \, 31  &    1.5  & -57.7  &  42.7  &  10.86 & 1.01 & 0.02  &  16.5  &47 &   78.46$\pm$ 0.21 &  80.60  \\
\object{2MASS~J22373980-2628544} & \object{1076} & 22 \, 37 \, 39.8 & -26 \, 28 \, 54  &   26.4  & -60.2  &  48.6  &  10.35 & 1.04 & 0.02  &  14.7  &49 &   15.13$\pm$ 0.22 &  64.33  \\
\object{2MASS~J22392246-2508120} & \object{1077} & 22 \, 39 \, 22.5 & -25 \, 08 \, 12  &   29.2  & -60.4  &  49.4  &   9.99 & 1.10 & 0.02  &  15.7  &18 &  -16.66$\pm$ 0.46 &  37.04  \\
\object{2MASS~J22442231-3247156} & \object{1078} & 22 \, 44 \, 22.3 & -32 \, 47 \, 16  &   13.5  & -62.0  &  48.1  &  11.21 & 1.03 & 0.01  &  21.5  &27 &   22.16$\pm$ 0.11 &  45.52  \\
\object{2MASS~J22561212-2045555} & \object{1080} & 22 \, 56 \, 12.1 & -20 \, 45 \, 56  &   40.3  & -63.0  &  54.4  &  11.24 & 0.99 & 0.03  &  18.6  &70 &  -50.66$\pm$ 0.16 &  14.34  \\
\object{2MASS~J23194353-1546105} & \object{1082} & 23 \, 19 \, 43.5 & -15 \, 46 \, 11  &   56.3  & -65.9  &  61.4  &  11.24 & 1.03 & 0.03  &  21.8  &63 &  -78.77$\pm$ 0.39 & -4.31   \\
\object{2MASS~J23212651-2426543} & \object{1083} & 23 \, 21 \, 26.5 & -24 \, 26 \, 54  &   35.4  & -69.6  &  58.6  &  10.50 & 1.14 & 0.02  &  24.0  &14 &  -19.21$\pm$ 0.16 &  23.63  \\
\object{2MASS~J23241474-2750339} & \object{1084} & 23 \, 24 \, 14.7 & -27 \, 50 \, 34  &   25.8  & -70.7  &  57.9  &  11.46 & 1.04 & 0.02  &  24.9  &34 &   -1.30$\pm$ 0.39 &  28.15  \\
\object{2MASS~J23262366-2500371} & \object{1085} & 23 \, 26 \, 23.7 & -25 \, 00 \, 37  &   34.5  & -70.8  &  59.4  &  10.78 & 1.00 & 0.02  &  15.7  &52 &  -20.79$\pm$ 0.14 &  18.25  \\
\object{2MASS~J23293008-2458096} & \object{1086} & 23 \, 29 \, 30.1 & -24 \, 58 \, 10  &   35.0  & -71.5  &  60.1  &  10.46 & 1.10 & 0.02  &  20.1  &15 &  -33.59$\pm$ 0.32 &  4.34   \\
\object{2MASS~J23295478-2051034} & \object{1087} & 23 \, 29 \, 54.8 & -20 \, 51 \, 03  &   47.2  & -70.4  &  61.8  &  10.59 & 1.00 & 0.04  &  14.4  &53 &  -42.89$\pm$ 0.10 &  9.67   \\
\object{2MASS~J23303457-1607474} & \object{1088} & 23 \, 30 \, 34.6 & -16 \, 07 \, 47  &   59.2  & -68.3  &  63.7  &  11.30 & 0.98 & 0.03  &  17.9  &71 &  -64.25$\pm$ 0.25 &  4.63   \\
\object{2MASS~J23430789-2358264} & \object{1089} & 23 \, 43 \, 07.9 & -23 \, 58 \, 27  &   40.7  & -74.3  &  63.4  &  11.35 & 0.97 & 0.02  &  17.8  &39 &  -29.11$\pm$ 0.08 &  6.93   \\
\object{2MASS~J23454168-2644555} & \object{1090} & 23 \, 45 \, 41.7 & -26 \, 44 \, 56  &   30.7  & -75.4  &  62.8  &  10.38 & 1.02 & 0.02  &  14.0  &48 &  -14.45$\pm$ 0.15 &  10.58  \\
\object{2MASS~J23503612-2002156} & \object{1091} & 23 \, 50 \, 36.1 & -20 \, 02 \, 16  &   56.7  & -74.4  &  66.5  &  10.58 & 1.07 & 0.02  &  18.4  &33 &  -55.91$\pm$ 0.30 & -9.18   \\
\object{2MASS~J23531941-2050407} & \object{1092} & 23 \, 53 \, 19.4 & -20 \, 50 \, 41  &   55.2  & -75.3  &  66.8  &  11.61 & 0.97 & 0.02  &  20.1  &55 &  -52.74$\pm$ 0.08 & -9.87   \\
\object{2MASS~J23563742-2347116} & \object{1093} & 23 \, 56 \, 37.4 & -23 \, 47 \, 12  &   45.0  & -77.2  &  66.3  &  11.46 & 0.99 & 0.02  &  20.8  &59 &  -41.64$\pm$ 0.20 & -10.71  \\
\object{2MASS~J23565416-1906094} & \object{1094} & 23 \, 56 \, 54.2 & -19 \, 06 \, 09  &   62.7  & -75.1  &  68.3  &  11.64 & 1.03 & 0.02  &  25.6  &49 &  -61.81$\pm$ 0.19 & -14.50  \\
\\
\hline
\end{tabular}
}
\\
\end{center}
\end{table*}

\section{Introduction}

Dwarfs are the most common type of galaxies in the universe.  Several  dwarf
satellites are usually associated with giant galaxies and, in the commonly
accepted scenario \citep{wr}, giant galaxies actually emerge out of the
hierarchical assembly of dwarfs. In this respect, dwarf galaxies are considered
to be the building blocks of the universe. 

The Sagittarius dwarf spheroidal galaxy (\citealt{s1}, \object{Sgr dSph}) is currently
disrupting into the Milky Way (MW) and its discovery historically represented
one of the first clear confirmations on a local framework of the hierarchical
merging paradigm. Nevertheless, the chemical analysis of stars in the MW
satellites and in Sgr itself seems to seriously challenge this evolutionary
scheme (see \citealt{venn} for a review). In fact,  it turned out that red giant
stars in local dwarfs have chemical patterns, in particular in the 
[$\alpha$/Fe] abundance ratios, that are not compatible with those of the
galactic halo  \citep[but see][for a possible solution to this problem]{rob}. 
However, the idea that dwarfs may have contributed the halo with stars even
significantly different from their present stellar population is now under
investigation \citep{carina,chou}. 

Tagging accreted components and analyzing their chemical composition is very
important for our comprehension of the Milky Way formation.  Some streams were
identified in the galaxy without any association with a core remnant. Therefore,
they could represent the residuals of ancient accretions \citep{virgo,orphan}, 
and their chemical signature might be very informative as well.

In this framework, Sgr plays a special role. It presents a very significant core
remnant (30$^\circ$ tidal radius), and its giant tidal streams (henceforth, the
Stream, for brevity), now identified all over the sky (\citealt{maje03},
hereafter M03), indicate that the disruption process is still ongoing. Hence,
Sgr is a MW satellite for which a complete reconstruction of the star formation
history is possible, combining core and stream information. As such, it will be
possible to understand if Sgr is actually a building block of the galactic halo
or not. Deep color magnitude diagrams (e.g., \citealt{marconi, sdgs2, ls00,
monaco02, conundrum}) and abundances derived from  high resolution spectra
\citep{B00,mc2,boni04,mc1,monaco05,luca} provided a fresh wealth of information
about the star formation history (SFH) of the stellar populations present in the
Sgr core over the years. Information about the Stream is now also accumulating
(M03, \citealt{maje04,law,belo,chou,bella}).

This is the first paper of a series devoted to the study of the Sgr Stream. 
Here, we present radial velocities for 67 red giant branch (RGB) stars belonging
to the  Stream and high resolution chemical abundances (Fe, Mg, and Ca) for 12
of these stars. The paper is organized as follows: in \S2 we describe the target
selection procedure. The observational dataset and the applied data reduction
procedure are discussed in \S3. In \S4 we describe the procedure for radial
velocity measurements and discuss the obtained results. In \S5 we present a
comparison between radial velocity obtained here and in \citet[][hereafter
M04]{maje04} for a sample of stars belonging to the Sgr trailing tail and common
to the two studies. In \S6 we present chemical abundances obtained for 12 stars
lying in two different spots of the Stream. In \S7 we discuss our findings. In
\S8 we briefly summarize the obtained results.

\section{Target selection}

Data were obtained using three different high resolution facilities. A
sub-sample of the M04 stars was observed with UVES (46 stars). The remaining
stars were selected from  the  2MASS~\footnote{See
http://www.ipac.caltech.edu/2mass.} catalog employing the M03 procedure, which
has already been proven to be a powerful tool to pick-out candidate stream stars
(see also M04). 

Reddening estimates were obtained through the \citet{cobe} reddening maps. 
Distances of the target stars were derived through photometric parallax,
following M03, but  adopting (m-M)$_0$= 17.10 as the distance modulus of the Sgr
core \citep{monaco04}, instead of 16.90.   Cartesian coordinates and longitudes
in the Sgr orbital plane were derived following M03 (see M03 for definitions and
details). Coordinates, magnitudes, and derived distances of the program stars
are listed in Table~\ref{coord}. Parameters for UVES stars are taken directly
from Table~3 in M04.

\begin{figure}
\centering
\resizebox{\hsize}{!}{\includegraphics[clip=true]{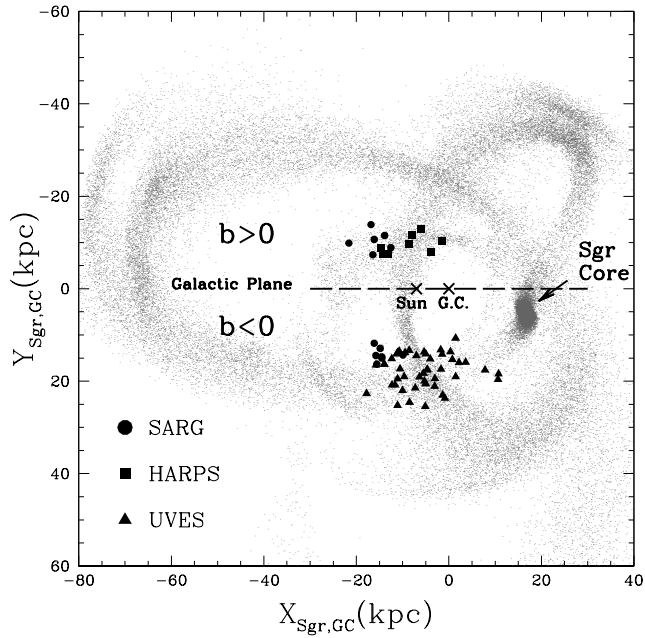}}
\caption{
Program stars positions in the Cartesian galactocentric Sgr plane (filled
symbols). Different symbols correspond to stars observed with different
spectrographs.  A model of the Sgr disruption (grey dots) is also plotted.  The
galactic plane (dashed line) and the position of the Sun, of the galactic
center, and of the Sgr main body are also marked for reference. 
}
\label{xysgr}
\end{figure}

In Fig.~\ref{xysgr} we plot the position of the target stars (big solid markers)
in Cartesian coordinates of the (galactocentric) Sgr orbital plane. Different
symbols correspond to stars observed with different facilities. The \citet{law}
model of Sgr destruction (for a spherical galactic potential) is also plotted
for reference. 

\section{Observations and data reduction}

A total of 13 stars were observed  between August 30, 2004, and January 24,
2005, using the SARG spectrograph mounted on the Telescopio Nazionale Galileo
(TNG) telescope at La Palma. We used the  1\farcs{6} slit, which provides a
resolution of R=29000. The chosen setup used the yellow cross-disperser, which
covers   approximately the 462--792~nm spectral range. Data reduction (bias
subtraction,  division by flat field, lambda calibration, background
subtraction, and extraction) was performed within the
ESO-MIDAS\footnote{ESO-MIDAS is the acronym for the European Southern
Observatory Munich Image Data Analysis System, developed and maintained by the
European Southern Observatory. http://www.eso.org/projects/esomidas/.} echelle
context. 

During a technical-time slot on the nights of June 3 and 4, 2006, we observed 8
supplementary stars  with the HARPS facility mounted at the 3.6m telescope in La
Silla.  The standard high resolution HARPS mode (R=110000, 380--690~nm spectral
range) was employed. Stars were observed for an integration time ranging from
800s (\#\object{465}) to 1200s (all the others).  Additional HARPS observations were
obtained for stars \#\object{1006}, \#\object{1022}, and \#\object{1083} in July 2006, with 30 min
exposures.  The June 29, 2006 star \#\object{1006} was observed for one hour
integration time. 

\begin{figure}
\centering
\resizebox{\hsize}{!}{\includegraphics[clip=true]{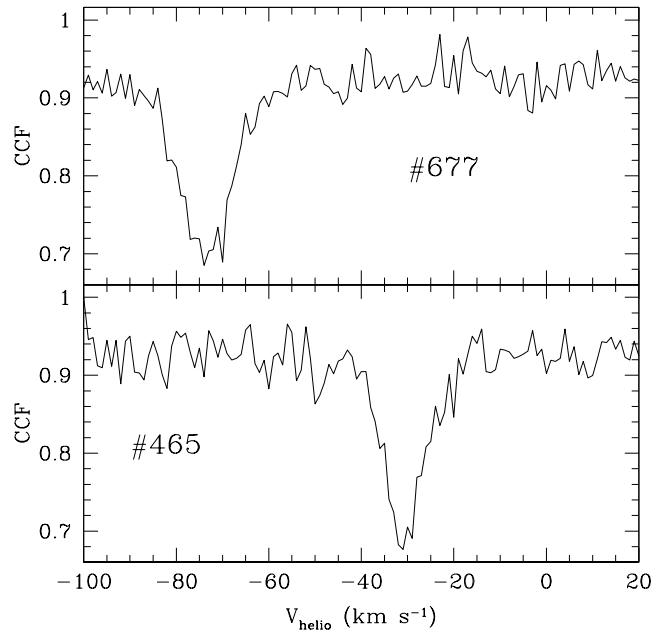}}
\caption{Cross correlation function (CCF) of the two lowest signal-to-noise 
HARPS spectra with the G2 dwarf template. The heliocentric star
radial velocity reported in Table~\ref{coord} is obtained through a 
Gaussian fit to the observed peak. Contrary to usual conventions, the   
star radial velocity is found by the HARPS pipeline as a minimum in the 
CCF.}
\label{ccf}
\end{figure}

Data were reduced through the online automatic pipeline installed on the WHALDRS
workstation at the 3.6m control room. The final output of the HARPS pipeline is
extracted spectra that are completely reduced (bias-subtraction,  cosmic rays
filtering, flat-field, and wavelength calibration), and the star radial velocity
as measured by a cross correlation function on the bidimensional echelle
spectrum with a template G2 dwarf\footnote{To date, the G2 dwarf is the only
template available for cross correlation using the HARPS pipeline. However,
using a not yet released M4 mask, we found a  $\sim$100~m~s$^{-1}$ radial
velocity difference in a test made on star \#\object{459992}.} mask. The extreme
stability of the HARPS facility secures accurate radial velocity measures even
with very low signal-to-noise spectra. In Fig.~\ref{ccf} we plot the cross
correlation function obtained for the two lowest S/N spectra obtained. The
signal corresponding to the star radial velocity is clearly evident.

UVES spectra for 46 stars were obtained between June 18 and September 16, 2005.
Stars were observed with the standard setting DIC~390+580~nm, which covers the
spectral range 328--456~nm and 480--680~nm, with the Blue and Red arms,
respectively.  We employed a 2$\times2$ CCD binning and a slit width of
1\farcs{2}, which provide a resolution of about 35000$\div$40000. Data were
reduced using the UVES ESO-MIDAS pipeline. 

\section{Radial velocities}

Radial velocities (RV) of star in the SARG and UVES samples are obtained by
cross correlation with a synthetic spectrum using the {\it fxcor} task inside
the IRAF\footnote{IRAF is distributed by the National Optical Astronomy
Observatories, which is operated by the association of Universities for Research
in Astronomy, Inc., under contract with the National Science Foundation.} suite.
The synthetic spectrum was calculated employing the SYNTHE code
\citep{k93,sbordone04} and a set of atmospheric parameters (temperature;
gravity; metallicity=3900; 1.0; -0.5) similar to those of all the observed stars
(see, e.g., Table~\ref{PA}). Concerning the HARPS spectra,  the formal photon noise
induced radial velocity error is in the worst case 11~m~s$^{-1}$. A conservative
200~m~s$^{-1}$ uncertainty is assumed. In Table~\ref{coord} we report the
measured radial velocities (heliocentric and in the galactic standard of
rest\footnote{A local standard of rest rotation velocity of 220~\kms and a
peculiar motion of (u,v,w)=(-9,12,7)~\kms are adopted for the Sun, for
consistency with M04.}), as well as the signal-to-noise ratio of the spectra.
Radial velocities are obtained with a precision, generally, better than
0.5~km~s$^{-1}$. For the HARPS and UVES spectra, we only present radial
velocities, here. 

\begin{figure}
\centering
\resizebox{\hsize}{!}{\includegraphics[clip=true, angle=-90]{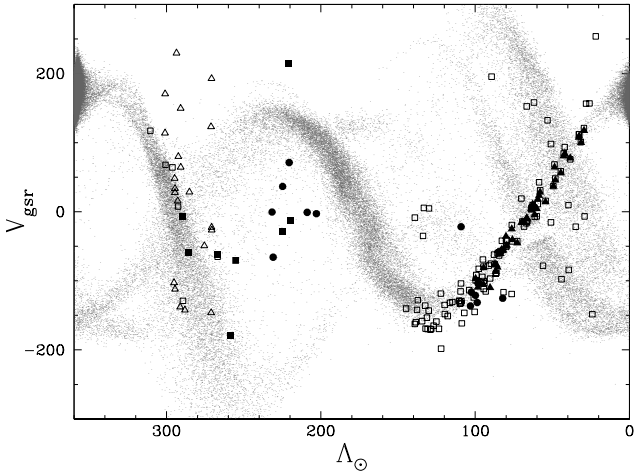}}
\caption{
Galactic standard of rest radial velocities of the program stars (filled
symbols) as a function of the  longitude of the Sgr orbital plane. Different
symbols correspond to stars observed with different spectrographs (circles,
squares, and triangles for SARG, HARPS, and UVES data, respectively).  Stars
studied by M04 (empty squares) and by \cite{spaghe} (open triangles) are also
plotted together with a model of the Sgr disruption (grey dots).
}
\label{lvr}
\end{figure}

In Fig.~\ref{lvr} we plot the program stars RVs (in the galactic standard of
rest, v$_{gsr}$)  as a function of the Sgr longitude scale ($\Lambda_\odot$)
along the orbital plane.  We also plot M04 (for distances larger than 13~kpc)
and \citet{spaghe} data superposed to the Sgr destruction model already used in
Fig.~\ref{xysgr}. Stars in the UVES sample describe a characteristic trend of
decreasing v$_{gsr}$ with increasing $\Lambda_\odot$ along the Sgr trailing
tail, as already discussed by M04. The same trend is also followed by SARG stars
at similar stream positions. Referring to Fig.~\ref{xysgr}, at positive galactic
latitude (b$>$0 or Y$_{Sgr,GC}<$0) all the SARG and 3 among the HARPS stars lie
on a well defined branch of stream (X$_{Sgr,GC}<$-10~kpc). The large dispersion
shown by this group in Fig.~\ref{lvr} ($\Lambda_\odot<230^\circ$) is predicted
to some extent by the model, and more data are mandatory to constrain the radial
velocity pattern of this part of the stream. The remaining part of the HARPS
stars lie in a region where different branches of the Sgr Stream overlap  
(X$_{Sgr,GC}>$-10~kpc in Fig.~\ref{xysgr}).  Their radial velocities nicely fit
with the trend predicted by the model for the Sgr leading tail and confirmed by
\citet{spaghe} (Fig.~\ref{lvr}) and \citet{law} data. However, especially for
the three stars at $230^\circ<\Lambda_\odot<280^\circ$, some ambiguity still
holds.

\section{The UVES sample: comparison with Majewski et al.~(2004)}

\subsection{Sanity check and possible binary stars}

\begin{figure}
\centering
\resizebox{\hsize}{!}{\includegraphics[clip=true,angle=-90]{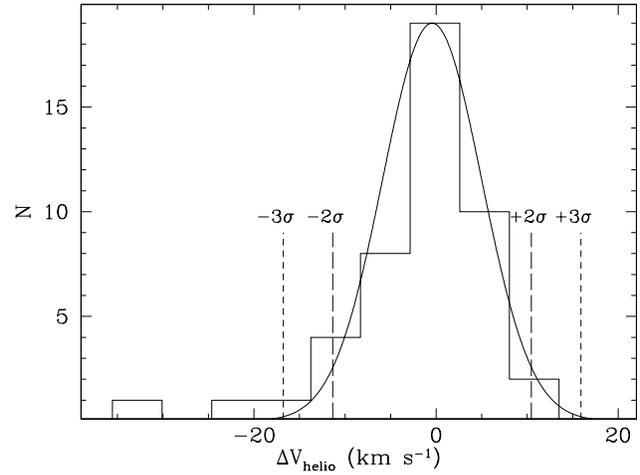}}
\caption{
Distribution of the difference between RVs measured here and in M04.
The best-fit Gaussian curve ($\sigma$=5.45~km~s$^{-1}$) is also plotted.
Long dashed and short dashed lines mark 2$\sigma$ and 3$\sigma$ levels.
}
\label{dv}
\end{figure}

UVES stars were selected out of the M04 sample and trace 70 degrees of the Sgr
trailing tail, in the range $30^\circ<\Lambda_\odot<100^\circ$. Note that
$\Lambda_\odot$=0 at the Sgr core. Figure~\ref{dv} shows the distribution of the
differences between the heliocentric radial velocity measured here and in M04
(see Table~\ref{coord}). After removing star \#\object{1006} (which shows a remarkably
large velocity difference: -35.4~km~s$^{-1}$), the distribution is well
represented by a Gaussian distribution centered at -0.44~km~s$^{-1}$ and having
a $\sigma$ of 5.45~km~s$^{-1}$. Hence, there is no zero point difference between
the two sets of measures and, given the high accuracy of the UVES velocities,
the dispersion of the distribution nicely confirms the 5.3~km~s$^{-1}$  quoted
by M04 as random errors.

It is noteworthy that 3 stars lie over the 3$\sigma$ limit ($>$16~\kms of RV
variation, Fig.~\ref{dv}). A possible reason for the detected RV difference is
that these stars are in fact binary systems, observed at different orbital
phases. Time series of RV measures are clearly needed to assess this hypothesis
on a firm basis. Between June and July 2006, additional HARPS data was obtained
for these stars. In Table~\ref{binary} we report a summary of the RVs measured
for stars \#\object{1006}, \#\object{1022}, and \#\object{1083}. Support of the
binary hypothesis is provided by this new data to star \#\object{1006} and, to
some extent, also to \#\object{1022},  while no significant RV  variation
between the UVES and HARPS measures was obtained for star \#\object{1083}. In
any case, considering the 3 outliers as genuine binaries, a preliminary lower
limit for the Sgr binary fraction of $\sim$6\% is derived.

\begin{table}
\caption{Heliocentric radial velocity at different dates for the three suspected
binaries. Brackets besides dates acknowledge UVES (U), HARPS (H), or M04 
measures. The last column reports the signal-to-noise ratio of the HARPS and UVES
spectra or the cross-correlation quality index for M04 data.}
\label{binary}
\begin{center}
\begin{tabular}{l|rrr|c}
\hline
   &  \multicolumn{3}{c|}{v~(\kms)} &\\
\hline
\multicolumn{1}{c|}{Date} & \object{1006} & \object{1022} & \object{1083} & Q\\
\hline
2002-07-15 (M04) &&&-0.1&7 \\
2002-07-30 (M04) &&-82.8&&5 \\
2002-07-31 (M04) &-6.5&&&7 \\
\hline
&&&& S/N\\
\hline
2005-07-19 (U)   &&& -19.21	&14\\
2005-07-20 (U) 	 &-41.90&&	&42\\
2005-09-13 (U)   &&-102.48&	&18\\
2006-06-29 (H)   &-46.65 &&	&10\\
2006-07-14 (H)   &&-105.14 &	&9\\
2006-07-15 (H)   &&&-19.71 &8\\
2006-07-17 (H)   &-47.33&& &13\\
2006-07-19 (H)   &-47.25&-104.64& &12, 11\\
\hline
\end{tabular}
\\
\end{center}
\end{table}

\subsection {The velocity dispersion of the Sgr trailing tail}

\begin{figure}
\centering
\resizebox{\hsize}{!}{\includegraphics[clip=true]{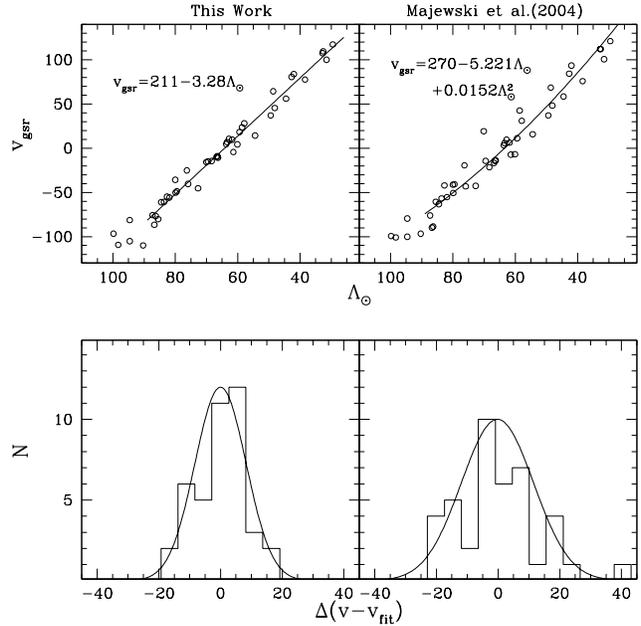}}
\caption{
Upper panels: measured radial velocities as a function of the longitude of the
Sgr orbital plane for stars in the UVES sample. The left panel shows measures
presented here; the right panel shows M04 measures. The fit to the observed
distributions are also plotted. Lower panels: distribution of the differences
between the fit and the actual RV for the two set of measures. 
}
\label{disp}
\end{figure}

In the upper panels of Fig.~\ref{disp}, we plot the v$_{gsr}$ as a function of 
$\Lambda_\odot$ for stars in the UVES sample. The left panel shows our measures,
and the right panel M04 RVs. Continuous lines show a least-squares fit and a
polynomial fit (M04) to the trend, in the former and latter cases, respectively.
The fits hold up to $\Lambda_\odot<$90$^\circ$, where the increase of the
velocity dispersion is evident (see M04).

Lower panels show the distribution of differences between the actual RV and the
fit. M04 data (right panel) is well fitted by a Gaussian curve having a
dispersion of 11.8~\kms, once star \#\object{1006} (which lies more than 3$\sigma$ away
from the mean) is removed. M04 used a $\sigma$=11.7~\kms Gaussian curve to fit
the observed distribution. Hence, the stars we observed are representative of
the more populous M04 sample. Note also that M04 used 45 stars to evaluate the
stream velocity dispersion, a number not so different from the 40 objects we use
here.

The left lower panel shows residuals of our measures with respect to the fit.
The distribution is fitted by a Gaussian of $\sigma$=8.3$\pm$0.9~\kms (without
rejecting any star, i.e., using 41 stars) while M04 obtained an intrinsic stream
dispersion of $\sigma$=10.4$\pm$1.3~\kms, after removing the random errors
($\sim$5.3~\kms). The two values are in agreement, within the errors.  However,
the above results suggest that this part of the trailing tail is colder than
what was estimated by M04 with low resolution spectroscopy, and it also appears
colder than the  Sgr core (11.17~\kms and 11.4~\kms in \citealt{monaco05},
hereafter M05, and \citealt{s2}, respectively). Nonetheless, the external
regions of the Sgr main body may present velocity dispersions more similar to
what observed in this portion of the Stream \citep[see also][]{s2}.

Our results suggest that to  properly characterize dynamical structures these
cold (e.g., streams in and outside the halo, dwarf galaxies velocity
dispersions), high resolution data are really useful, if not mandatory. It
should be also kept in mind that a sizable population of binaries could  (and
indeed should) be present in Sgr. However, the increase of the measured velocity
dispersions of a dwarf galaxy due to the presence of a binary population should
be considered at most marginal \citep[][]{har,ols}.

At $\Lambda_\odot>$90, we confirm the M04 claim of a rise in the stream velocity
dispersion. However, with just 5 stream stars  no meaningful comparison with M04
can be done. Note also that a colder velocity dispersion in the stream of a
disrupting system with respect to the core remnant is expected on the basis of
the conservation of phase-space density \citep[see][]{he}. The stream velocity
dispersion should actually decrease as a function of time (as $\frac{1}{t}$).

\section{Chemical analysis}\label{atm}

Looking at Fig.~\ref{xysgr}, it is easy to realize that SARG stars  lie at the
most extreme Stream positions, among the program stars. They sample two
different wraps of Stream and in each of them a part of the Stream significantly
distant from the Sgr core.  Therefore, SARG stars are ideal to spot the basic
chemical characteristics of the Sgr Stream. Here we present Fe, Mg, and Ca
abundances for the stars in the SARG sample.

\begin{figure*}
\centering
\resizebox{\hsize}{!}{\includegraphics[height=6.5cm, width=7cm]{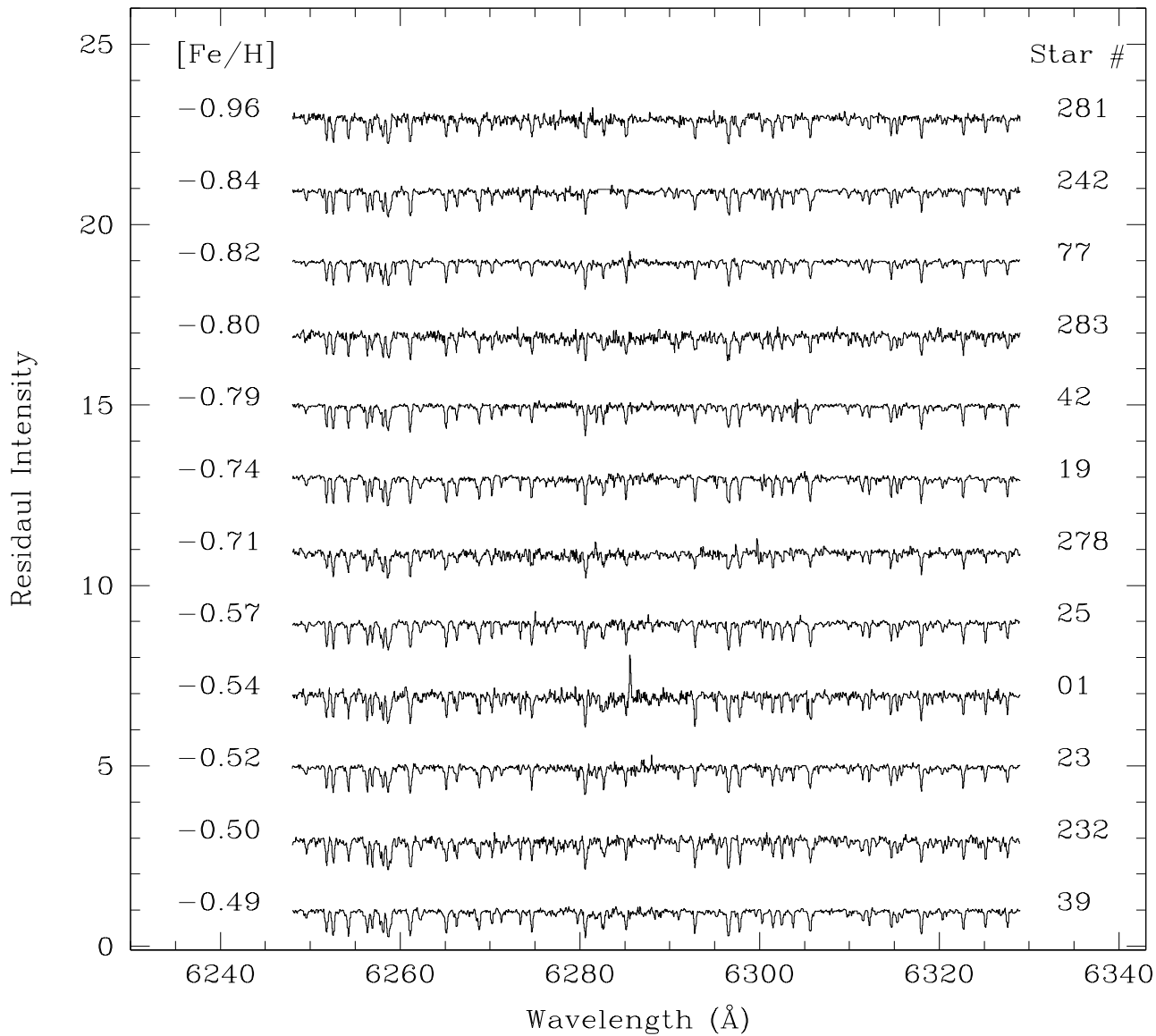}}
\caption{
Sample of the SARG spectra of the 12 stream stars for which the chemical 
analysis was performed. Labels on the right
denote the star number, those on the left the measured [Fe/H].
}
\label{spettri}
\end{figure*}

In more detail, stars from \#\object{1} to \#\object{77} of Table~\ref{coord} belong to the
trailing tail (b$<$0 or Y$_{GC,Sgr}>$0, Fig.~\ref{xysgr}) at more than 80
degrees from the Sgr core.  Stars from \#\object{232} to \#\object{283} (b$>$0 or Y$_{GC,Sgr}<$0),
lie above the galactic plane and probably belong to a more ancient branch of the
stream. Star \#\object{260} has a too low S/N ratio  (see Table~\ref{coord}) to allow a
reliable chemical analysis and is,  therefore, dropped in the following
discussion. In Fig.~\ref{spettri} we plot a sample of the SARG spectra of the 12
stars for which the chemical analysis was performed.

Stars in the UVES sample have cooler temperatures with respect to SARG
stars.  As such, the great majority of them present deep titanium oxide bands
(TiO, see Fig.~4 in M05), which strongly complicates the chemical analysis.  
TiO bands depress the continuum, and a reliable estimate of the continuum level
is crucial for robust equivalent width measurements. Thus, the derivation of elemental
abundances for such cool stars represents a significant challenge. A few groups
are actively investigating methods to derive trustworthy abundances for M stars
by the simultaneous comprehensive synthesis of selected spectral regions roughly
in the range 7000\AA$<\lambda<$9000\AA  \citep[see][]{va98,bean}. For this
reason  and in spite of the high quality of the data, the analysis of stars in
the UVES sample will be presented in a forthcoming contribution.

\subsection{Atmospheric parameters and chemical abundances}

Dereddened (J-K) colors were used together with the \citet{alonso} calibration
to derive the effective temperature (T$_{eff}$) of the program stars. Stars
share very similar colors ((J-K)$_0$=0.93$\div$0.97, see Table~\ref{coord}),
which turn into a quite tight range of temperature, namely T$_{\rm
eff}$=3831$\div$3936~K.  Note that effective temperatures derived with this
procedure, however, appear on average roughly 2\% hotter (i.e., +76~K) than the
calibration scale adopted in M05\footnote{In M05, T$_{eff}$ were derived for
stars in the Sgr core from optical colors. The quoted 2\% of difference in the
temperature scale was estimated comparing M05 stars temperatures as derived from
optical and infrared colors.}. We eventually adopted T$_{eff}$=3900~K for all of
our stars, assuming a $\pm$100~K uncertainty. However, note that (i) the assumed
T$_{eff}$ obtain excitation equilibrium of the neutral iron lines (Fe~I) in all
but two (\#\object{242} and \#\object{42}) of the stream stars,  and that (ii) 76~K of difference
in the temperature scale  do not induce any sensible change  in the derived
abundances, as can be seen from Table~\ref{errors} and Table~4 in M05.

All the targets were photometrically classified as M-Giants (see M03).  However,
photometric classification is always tentative and should be spectroscopically
confirmed. Stars are classified as M-type on the basis of the presence of
titanium oxide (TiO) bands in their spectra. Indeed, the SARG spectra do not
present TiO bands, as somewhat expected from their not exceedingly low
temperatures (see also M05). Thus, chemical abundances are safely derived from
spectral lines equivalent widths (EW) provided a proper model atmosphere is
employed.

After correcting for their distance and reddening, gravity should 
be derived for target stars by the relevant fundamental relationship:
\begin{equation} 
\log g = 4\,\log T_{\rm eff} -\log L_* + \log M_* +{\rm const}, 
\end{equation} 
where ${\rm const} = \log (4\pi G\sigma) = -10.32$ and $M_*$ and $L_*$ are the
stellar mass and luminosity. However, given the obvious uncertainty in the
definition of both the stellar mass and the bolometric correction at such low
temperatures, only a safe physical range can be identified for $\log g$, relying
on a collection of isochrones \citep[see, e.g., Fig.~11 in][]{bertone}. 
Comparing with the \citet{leo} isochrones in the K vs. (J-K) plane, we
derived $\log g = 0.9\pm 0.5$~dex as a realistic estimate of the representative
surface gravity and its allowed range, for all the targets.

To derive the chemical abundances, we firstly calculated a model atmosphere with
T$_{eff}$=3900~K, log~g=0.9, [M/H]=-0.5, and the Opacity Distribution  Functions
of \citet{1993KurCD}. Secondly, we measured EWs on the spectra for a selected
sample of Fe, Mg, and Ca lines using the standard IRAF task {\it splot}.
Finally, abundances were derived from the measured EWs using the calculated
model atmosphere within the WIDTH code.  The GNU-Linux ported version
\citep{sbordone04} of both the WIDTH and ATLAS codes \citep{k93} were
employed.   Microturbulent velocities ($\xi$) for each star were determined
minimizing the dependence of the iron abundance from the EW. 

\begin{table}
\caption{Atmospheric parameters assumed for the program stars.}
\label{PA}
\begin{center}
\begin{tabular}{lrrrr}
\hline
\\
Star & $\rm T_{eff}$ &log g & $\xi$ & [M/H]  \\
\\
\hline
\\
\object{1}    &   3900 & 0.9 & 2.1 & $-0.5$\\
\object{19}   &   3900 & 0.9 & 2.1 & $-0.5$\\
\object{23}   &   3900 & 0.9 & 2.0 & $-0.5$\\
\object{25}   &   3900 & 0.9 & 2.0 & $-0.5$\\
\object{39}   &   3900 & 0.9 & 1.7 & $-0.5$\\
\object{42}   &   3900 & 0.9 & 2.1 & $-0.5$\\
\object{77}   &   3900 & 0.9 & 1.9 & $-0.5$\\
	\\
\object{232}  &   3900 & 0.9 & 2.1 & $-0.5$\\
\object{242}  &   3900 & 0.9 & 2.2 & $-0.5$\\
\object{278}  &   3900 & 0.9 & 2.0 & $-0.5$\\
\object{281}  &   3900 & 0.9 & 2.2 & $-0.5$\\
\object{283}  &   3900 & 0.9 & 2.0 & $-0.5$\\
\\
\hline
\end{tabular}
\\
\end{center}
\end{table}

The atmospheric parameters adopted for the program stars are reported in
Table~\ref{PA}. The Fe, Ca, and Mg line lists, as well as the adopted atomic
parameters and the measured EW, are reported in Table~\ref{abund1}. 
Table~\ref{abund1} also lists the abundance obtained for each line.  The mean
and standard deviation of such abundances can be found in Tables~\ref{abund0}
and \ref{abund} (as [X/H] abundances in the latter case) for each chemical
species together with the number of lines employed. The line scatter reported in
Tables~\ref{abund0} and~\ref{abund}  should be representative of the
statistical error arising  from the noise in the spectra and from uncertainties
in the measurement of the  equivalent widths\footnote{Under the assumption that
each line provides an independent  measure of the abundance, the error in the
mean abundances should be obtained by dividing the line scatter by $\sqrt{n}$
(where  $n$ is the number of measured lines). However, we consider the line
scatter reported in Tables~\ref{abund0} and~\ref{abund} (which is {\it not}
divided by $\sqrt{n}$) as a realistic  estimate of the error associated with
each abundance.}. In Table~\ref{errors}  we report the errors arising from the
uncertainties in  the atmospheric parameters in the case of star \#\object{19}, taken as
representative of the whole sample.

\begin{table}
\caption{Mean chemical abundances of the program stars. 
The number of lines used and the line scatter are also reported.}
\label{abund0}
\begin{center}
{\scriptsize
\begin{tabular}{llclclc}
\hline
\\
Star & A(Fe) & $n $ & A(Mg) & $n$ & A(Ca) & $n$\\
\hline
\\
Sun   & 7.51	      &    & 7.58	   &   & 6.35	       &  \\
 \object{1}    & 6.96$\pm$0.28 & 28 & 7.08$\pm$0.19 & 4 & 5.57$\pm$0.20 & 7\\
 \object{19}   & 6.77$\pm$0.22 & 36 & 6.80$\pm$0.25 & 4 & 5.38$\pm$0.19 & 8\\
 \object{23}   & 7.00$\pm$0.27 & 33 & 6.83$\pm$0.12 & 2 & 5.51$\pm$0.14 & 7\\
 \object{25}   & 6.94$\pm$0.22 & 31 & 7.15$\pm$0.25 & 4 & 5.52$\pm$0.14 & 7\\
 \object{39}   & 7.02$\pm$0.22 & 33 & 6.80$\pm$0.14 & 3 & 5.56$\pm$0.17 & 7\\
 \object{42}   & 6.70$\pm$0.22 & 32 & 6.93$\pm$0.09 & 3 & 5.41$\pm$0.25 & 8\\
 \object{77}   & 6.69$\pm$0.25 & 32 & 7.00$\pm$0.06 & 4 & 5.63$\pm$0.27 & 6\\
&&&&&&\\
\object{232}   & 7.01$\pm$0.20 & 25 & 7.51$\pm$0.20 & 4 & 5.94$\pm$0.23 & 7\\
\object{242}   & 6.67$\pm$0.29 & 27 & 7.05$\pm$0.11 & 3 & 5.49$\pm$0.15 & 7\\
\object{278}   & 6.80$\pm$0.24 & 17 & 6.89$\pm$0.14 & 3 & 5.54$\pm$0.16 & 7\\
\object{281}   & 6.54$\pm$0.18 & 25 & 6.83$\pm$0.19 & 4 & 5.14$\pm$0.17 & 7\\
\object{283}   & 6.71$\pm$0.20 & 26 & 6.94$\pm$0.15 & 4 & 5.41$\pm$0.23 & 5\\
\\
\hline 
\\
\multispan{3}{A(X)=log($\frac{X}{H}$)+12.00}
\\
\\
\end{tabular}
}
\end{center}
\end{table}

\begin{table}
\caption{Mean abundance ratios for the program star. For iron abundances, the 
line scatter is also reported.}
\label{abund}
\begin{center}
\begin{tabular}{lrrr}
\hline
\\
Star      & \multispan1{\hfill[Fe/H]\hfill}&   [Mg/Fe] & [Ca/Fe]\\ 
\\
\hline
\\
\object{1}  & -0.55$\pm$0.28 &  0.05 & -0.23\\
\object{19} & -0.74$\pm$0.22 & -0.04 & -0.23\\
\object{23} & -0.51$\pm$0.27 & -0.24 & -0.33\\
\object{25} & -0.57$\pm$0.22 &  0.14 & -0.26\\
\object{39} & -0.49$\pm$0.22 & -0.29 & -0.30\\
\object{42} & -0.81$\pm$0.22 &  0.16 & -0.13\\
\object{77} & -0.82$\pm$0.25 &  0.24 &  0.10\\
	&&&\\
\object{232}& -0.50$\pm$0.20 &  0.43 &  0.09\\
\object{242}& -0.84$\pm$0.29 &  0.31 & -0.02\\
\object{278}& -0.71$\pm$0.24 &  0.02 & -0.10\\
\object{281}& -0.97$\pm$0.18 &  0.22 & -0.24\\
\object{283}& -0.80$\pm$0.20 &  0.16 & -0.14\\
\\
\hline
\\
\multispan{3}{[X/Y]=log($\frac{X}{Y}$)-log($\frac{X}{Y}$)$_\odot$}
\\
\end{tabular}
\\
\end{center}
\end{table}

\begin{table}
\caption{Errors in the abundances of star \#\object{19} due to uncertainties
in the atmospheric parameters.}
\label{errors}
\begin{center}
\begin{tabular}{lrrrrrr}
\hline
\\
 & $\Delta$A(Fe) & $\Delta$A(Mg) &$\Delta$A(Ca)\\
\hline
\\
$\Delta\xi =  \pm 0.2$ \kms &  $^{-0.10}_{+ 0.12}$ & $\mp 0.04$ & 
$^{-0.11}_{+ 0.12}$\\
\\
$\rm \Delta T_{eff} = \pm 100$ K &$ ^{-0.01}_{+0.04}$ & $^{-0.01}_{+0.04}$ &
$^{+0.10}_{-0.09}$\\ 
\\
$\Delta \log g = \pm 0.50 $ & $^{+0.15}_{-0.14}$ & $^{+0.08}_{-0.07}$ &
$^{-0.01}_{+0.00}$\\
\\
\hline
\\
\end{tabular}
\end{center}
\end{table}

\begin{figure}
\centering
\resizebox{\hsize}{!}{\includegraphics[clip=true]{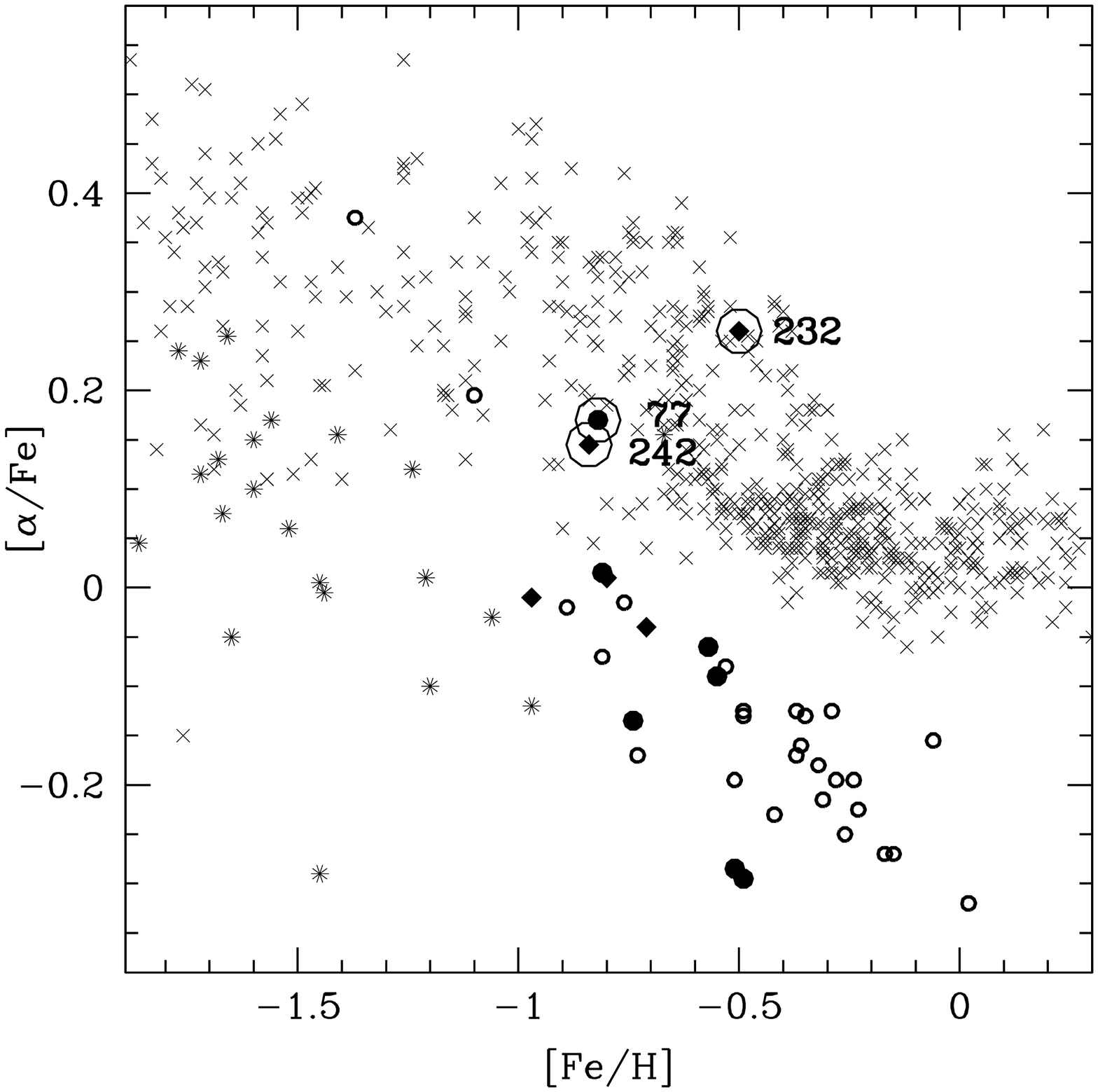}}
\caption{[$\alpha$/Fe]=$\frac{[Ca/Fe]+[Mg/Fe]}{2}$ as a function of the [Fe/H] 
for MW stars and local group galaxies (crosses and asterisks, respectively, from
Venn at al.~2004). Filled symbols refer to program stars (diamonds for the b$>$0
subsample) while Sgr main body 
stars are plotted as empty circles.}
\label{alphame}
\end{figure}

In Fig.~\ref{alphame} we plot the mean alpha element abundance ratio (defined as
$<[\alpha/Fe]>=\frac{[Mg/Fe]+[Ca/Fe]}{2}$ as in M05)  as a function of the
measured [Fe/H]. Chemical abundances of main body stars from M05\footnote{Only
the 15 stars   not showing TiO bands are plotted. See Table~1 in M05.} and
B04\footnote{The B04 abundances were  recomputed adopting the same temperature
scale and reddening adopted in M05. These variations in the input parameters
produced small (compatible with the quoted errors) changes in the derived
abundances \citep[see][]{luca}.}  are plotted. MW and Local group dwarfs stars
are plotted as well \citep[][]{venn}. Stream stars, clearly, follow the same
trend defined by the stars that are still bound to the core of Sgr.  Stars
belonging to the b$>$0 subsample are indicated.

Stars \#\object{77}, \#\object{232}, and \#\object{242} occupy a portion of plane dominated by MW stars in
Fig.~\ref{alphame}. Unlike \#\object{232}, stars \#\object{77} and \#\object{242} lie in a transition
region where their abundances are still compatible with the Sgr path. Star \#\object{77}
also has a relatively high RV (v$_{gsr}$=-21.1) compared with the mean stream
pattern ($\Lambda_\odot$=109.2, see Fig.~\ref{lvr} and Table~\ref{coord}). On
the other hand, \#\object{242} RV is similar to other SARG and HARPS stars lying at
similar stream longitudes (Fig.~\ref{lvr}). Hence, in the following analysis we
conservatively drop \#\object{77} and \#\object{232} as possible contaminating MW stars. We keep
star \#\object{1}, in spite of its slightly low RV, since its chemical composition
follows the Sgr pattern. The inclusion or exclusion of this star does not
substantially modify our conclusions.

\section{Discussion}

We presented RV for a sample of 67 stars belonging to the Sgr Stream.  Spectra
were obtained using 3 different high resolution facilities, namely SARG@TNG,
HARPS@3.6m, and UVES@VLT. Stars in the UVES sample (46 stars) trace 70$^\circ$
along the trailing tail and were already observed at low resolution by M04.   We
found a trailing tail velocity dispersion of 8.3$\pm$0.9~\kms, a value in
marginal agreement with M04 (10.4$\pm$1.3~\kms) and colder than the Sgr core
(\citealt{s2}, M05). The reader is referred to M04 for a discussion about the
implications of the velocity dispersion in the Stream for the lumpiness of the
galactic halo. We just recall here that a lumpy halo tends to heat coherent
streams.  However, the part of Stream we sample is populated by stars stripped
in relatively recent times and, therefore, is probably not very sensitive to the
lumpiness of the halo. We also presented Fe, Mg, and Ca abundances for 12 stars
observed with the SARG facility. Ten of them  are {\it bona fide} Sgr stream
members as of their chemical  abundances and RV (Figs.~\ref{alphame}
and~\ref{lvr}). Note, however, that any individual star can only be considered
as a probable member.

\begin{figure}
\centering
\resizebox{\hsize}{!}{\includegraphics[clip=true]{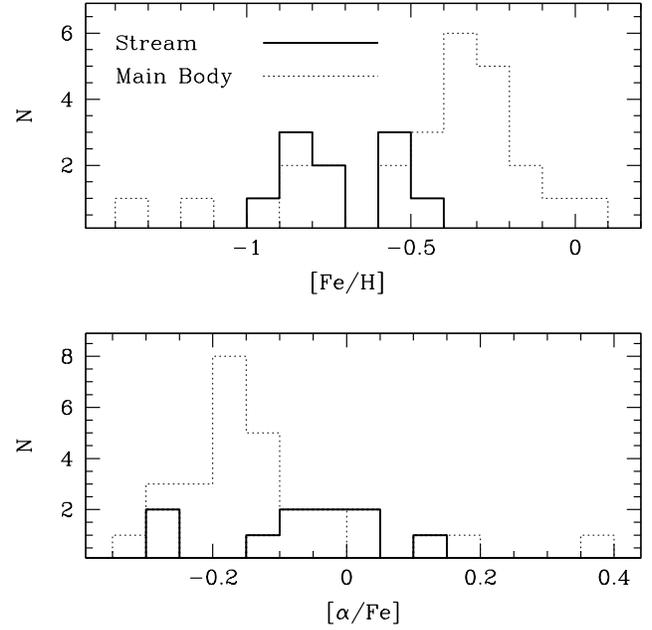}}
\caption{
Comparison between the metallicity (upper panel) and [$\alpha$/Fe] (lower panel)
distribution of main body (dotted histogram) and stream (continuous histogram)
stars.
}
\label{fedistr}
\end{figure}

In Fig.~\ref{fedistr} (upper panel) we compare the Sgr main body (dotted
histogram) and Stream (continuous histogram) metallicity distribution (MD).  We
point out that in M05, target stars were chosen in the infrared K vs. (J-K)
plane adopting the selection box of Fig.~1 in that paper. In the infrared plane,
in fact,  the upper Sgr RGB stands out very clearly from the contaminating MW
field (to compare with the optical plane, see Fig.~2 in M05). Indeed, such a
selection implies a bias toward metal rich stars, and, actually,  we provided a
thorough sampling of the Sgr dominant population \citep[][]{monaco02} at the Sgr
center (i.e., around the globular cluster M~54, whose RGB is roughly represented
by the bluer isochrone in Fig.2 of M05).

The existence of a metal rich dominant population in Sgr allowed M03 to  develop
his successful technique for tracing the Sgr streams all over the sky.  We used
such a technique here to select our targets. It is easy, looking at Fig.~1 (and
2) in M03, to realize that the M03 and M05 selection criterion are practically
the same. Note, that the mean temperature and gravity of the 15 stars analyzed
in M05  (the first 15 lines of Table~1 in M05\footnote{Note that stars marked
with an asterisk in Table~1 of M05 shows TiO bands and are not analyzed for
chemical abundance there.}) are 3975~K and 1.00 (with 177~K and 0.18 as standard
deviations, respectively) against the 3900~K and 0.9 adopted here for our stars.

B04 adopted a different selection function. Essentially, they selected fainter
stars, which have slightly larger gravities. The abundances derived in M05 and
B04 are compatible with each other within the errors \citep[see M05 and][for 
discussions]{luca}.

Main body stars show a well defined peak in the MD at $\sim$-0.35
($<$[Fe/H]$>$=-0.35$\pm$0.19, considering stars with [Fe/H]$>$-0.80)  and sample
a large metallicity range. Stream stars sample a smaller metallicity range, as
somewhat expected by the small number of stars analyzed. Clearly, the stream MD
is shifted toward lower metallicity ($<$[Fe/H]$>$=-0.70$\pm$0.16) compared to
the main body. A Kolmogorov-Smirnov test provides a probability $<$10$^{-3}$
that core and stream stars are extracted from the same parent population.  A
similar effect is also evident in  the distribution of core and stream stars
alpha element abundance ratio (lower panel).

\begin{figure}
\centering
\resizebox{\hsize}{!}{\includegraphics[clip=true]{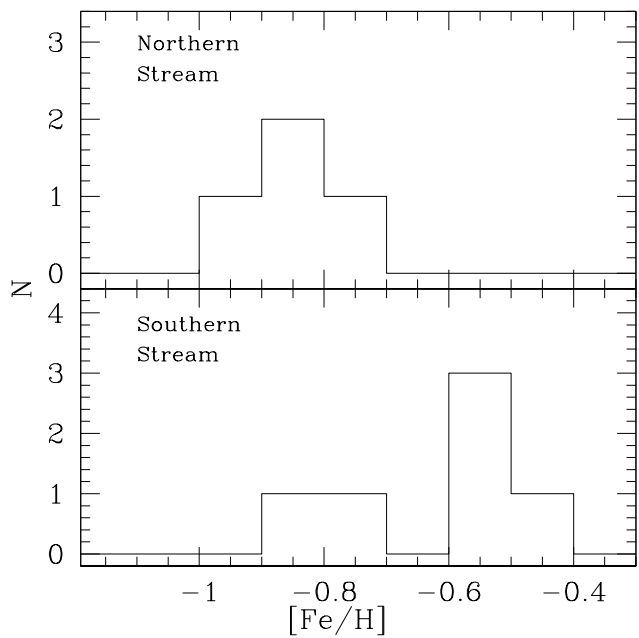}}
\caption{
Comparison between the metallicity distribution of stars in the Sgr trailing
tail (lower panel) and in a portion of stream above (upper panel) the galactic
plane. Stars in the upper panel sample a more ancient branch of the stream with
respect to stars in the lower panel (see Fig.\ref{xysgr}). 
}
\label{nordsud}
\end{figure}

SARG stars sample very different regions of the Stream (see Fig.~\ref{xysgr}).
The sub-sample at negative galactic latitude (stars from \#\object{1} to \#\object{42}; b$<$0 or
Y$_{Sgr,GC}>$0 in Fig.~\ref{xysgr}) belongs to the Sgr trailing tail in the
$80^\circ<\Lambda_\odot<100^\circ$ region. Hence, they were probably stripped
during the last Sgr orbit. The b$>$0 sub-sample (stars from \#\object{242} to \#\object{283})
traces, on the other hand, a more ancient episode of tidal stripping. In
particular, according to the \citet{law} model of Sgr disruption, they should
have been lost three or more orbits ago (i.e., $>$2-3~Gyr ago).  In
Fig.~\ref{nordsud} we plot the two sub-samples MD.  It is evident that the mean
metallicity of northern (b$>$0) stream stars is lower compared to southern ones:
$<$[Fe/H]$>$=-0.83$\pm$0.11 and $<$[Fe/H]$>$=-0.61$\pm$0.13, respectively. 
However, given the small number of stars in the two subsamples, this result has
to be considered tentative and must be confirmed by the analysis of a more
statistically significant sample of stars.

In summary, we found evidence of a more metal-poor MD in the Sgr Stream
compared to the main body. Moreover, stars stripped in ancient orbits appear
more metal-poor than stars lost in  recent passages.
Preliminary results pointing in the same direction were reported by
\citet{david} and M03. Recently, more definitive indications in this sense were
provided by \citet{chou} (hereafter C06) and \citet{bella}. 

In particular, C06 presented high resolution iron abundances for 56 M-giants
belonging to the Sgr leading tail. Thus, we sample a different and complementary
stream region.  C06 found a variation of about -0.7~dex in the mean iron content
from the core to the portion of stream they sample. Our results are
qualitatively in agreement with C06, although the variation we find is a bit
smaller. Several reasons can be responsible for such a difference (i.e., the
different stream portion sampled, their more populated sample, the different
analysis). C06 interpreted their results as a ``direct evidence that there can
be significant chemical differences between current dSph satellites and the bulk
of the stars they have contributed to the halo". Our results confirm this
statement.  Moreover, as we pointed out in M05, Sgr stars at [Fe/H]$<$-1 would
also have [$\alpha$/Fe] abundance ratios similar to MW stars (see
Fig.~\ref{alphame}), hence they may be eligible as contributors to the assembly
of the ``normal'' Galactic Halo (i.e., metal-poor and $\alpha$-enhanced).  
However,  \citet{luca} found under-solar or over-solar abundances for several
elements in 12 Sgr core stars. Even more interesting, they found a flat trend
over the range -0.9$<$[Fe/H]$<$0 in some of the anomalous abundance ratios such
as [Na/Fe], [Zn/Fe], and [Cu/Fe]. Detailed abundances of Stream stars,
especially of the most metal-poor ones, will be of the uttermost importance to
finally establish whether Sgr stars lost in ancient passages could have
significantly contributed to the standard stellar population of the galactic
halo or not.

Moreover, the galactic halo is populated by very old stars. Under the assumption
that dSph galaxies are dominated by an intermediate age population
(``Carina-like"),  \citet[][]{unavane} concluded  that no more than 10\% of the
whole halo stellar population may have originated from accretion \citep[but
see][]{carina}. Indeed, these authors predicted low [$\alpha$/Fe] abundance
ratios in the MW satellites as a result of their  inferred low star formation
rates  (see Fig~\ref{alphame}).

The Sgr stellar content is dominated in the main body by an intermediate age
\citep[][]{conundrum} population. Stream stars studied here follow the abundance
pattern of main body stars (Fig.~\ref{alphame}) and are moderately  more metal
poor than them (Fig.~\ref{fedistr}).  Hence, our targets likely have ages not
much older than stars in the Sgr dominant  population. Indeed, the SFH implied
by the chemistry of stars sampled here, do not differ significantly from that of
core stars: a prolonged period of star formation is needed to reach low
[$\alpha$/Fe] abundance ratios.  The C06 sample may eventually be made by older
stars, more similar to the  typical stellar content of the galactic halo.

However, neither our study nor the C06 one provide a fair sampling of the MD of
the Sgr stream. We both selected targets  using the 2MASS catalog. As such, the
target selection biases our samples toward metal rich stars.  As stated above,
the target selection box was actually shaped to enclose stars in the  Sgr
dominant population. Thus, the actual  MD of the Stream might eventually be
skewed to even lower metallicities and made by older stars, for any reasonable
age-metallicity relation \citep[see, e.g.,][]{ls00}. Note also that Sgr is known
to host a significant population  of old and metal-poor stars \citep[both in the
main body and the streams, see][and references therein]{rrstream,monaco03}.

Even SARG stars at b$>$0 could not have been stripped more then a few Gyr ago, a
time at which the Sgr star formation was already completed. Therefore, we agree
with the C06 conclusion that the difference in the core and stream MD witness a
metallicity gradient inside the former Sgr (see, for instance,
\citealt{sdgs2,ls00}, and references therein). Chemical abundances in the
outskirts of the Sgr main body would be necessary to quantify metallicity
gradients inside Sgr \citep[see also][]{alard}. Stars in the trailing tail
(lower panel in Fig.~\ref{nordsud}) are only mildly more metal poor than core
ones (upper panel in Fig.~\ref{fedistr}). Hence, eventually, stars lying in the
outer Sgr core and in the trailing tail might not present any chemical
difference.

Indeed, the great majority of the MW satellites contain populations of old
stars, either dominant or not. It appears a general tendency of the most 
metal-poor populations in dSphs to be less concentrated with respect to the
other populations hosted \citep[][and references therein]{carina, tolstoy}. 
This characteristic might favor the preferential stripping of metal-poor stars
during tidal interactions between dSphs and the MW.

\section{Conclusions}

The Sgr SFH, its dynamical status and orbital evolution are constrained by the
stellar populations hosted both in the main body and in the tidal streams of
this disrupting galaxy. In this paper we presented radial velocities and
chemical abundances (Fe, Mg, Ca) for a sample of stars belonging to the Sgr
tidal streams.  In particular, we presented the first $\alpha$-element
abundances ever obtained for stars in the Sgr stream.  The main results obtained
can be summarized as follows:

\begin{itemize}
\item The velocity dispersion of the Sgr trailing tail (8.3~\kms) is 
significantly lower than in the main body (11.2~\kms).
\item Stream stars follow the same distinct trend described by stars in the Sgr
main body in the [Fe/H] vs. [$\alpha$/Fe] plane (Fig.~\ref{alphame}).
\item Sgr stars are, on average, more metal-poor in the Stream than in the 
core (Fig.~\ref{fedistr}).
\item Stars belonging to more ancient wraps of the Streams are more 
metal-poor (Fig.~\ref{nordsud}). This result was obtained comparing 
the MD of stars belonging to two different wraps of the Stream. However, given
the limited number of stars in the two subsamples (4 and 6), this latter 
result has to be considered tentative.
\end{itemize}

\begin{acknowledgements} 

We are grateful to A. Magazz\`u for his help in preparing the SARG observations.
We also thank the La~Silla SciOps department for their help in dealing with the 
HARPS spectra. Part of the data analysis has been performed using software
developed by P. Montegriffo at the INAF - Osservatorio Astronomico di Bologna.
The routines available at http://www.astro.virginia.edu/$\sim$srm4n/Sgr/  were
used to derive coordinates in the Sgr orbital plane.
We acknowledge support from the MIUR/PRIN 2004025729, the
INAF/PRIN05~CRA~1.06.08.02  and the INAF/PRIN05~CRA~1.06.08.03. PB also
acknowledges support   from EU contract MEXT-CT-2004-014265 (CIFIST).

\end{acknowledgements}

\bibliographystyle{aa}

\appendix

\section{Individual line data} 

The following tables report the  line list and adopted atomic parameters for
the program stars. The measured equivalent width and the corresponding
abundance obtained for each line are also reported.

\begin{table*}
\begin{center}
\caption{Line list and adopted atomic parameters for the program stars. The measured equivalent width and the
corresponding abundance obtained for each line are also reported.}
\label{abund1}
\scriptsize{
\begin{tabular}{rrrlrrrrrrrr}
\hline
\\
Ion  & $\lambda$ & log gf  & source of   & EW       &$\epsilon$ & EW       & $\epsilon$& EW (pm) & $\epsilon$ & EW      & $\epsilon$ \\
     & (nm)      &         & log gf      & (pm)     &           & (pm)     &           & (pm)    &            & (pm)    &            \\
     &           &         & (see notes) &  \object{1}       &           & \object{19}       &           & \object{23}      &            & \object{25}      &	     \\
\\
\hline
\\
Fe I & 585.5076  & -1.76   & FMW          & --     & --     & --    & --    &  5.09 & 7.283 & --    & -- \\	        
Fe I & 588.3817  & -1.36   & FMW          & --     & --     & 9.47  & 6.640 & 12.15 & 7.151 & 12.15 & 7.151	\\      
Fe I & 595.2718  & -1.44   & FMW          &12.89   & 7.324  & 9.27  & 6.719 & 12.70 & 7.360 &  7.62 & 6.491	\\      
Fe I & 602.7051  & -1.21   & FMW          &10.52   & 6.818  &10.78  & 6.861 &  9.28 & 6.661 & 11.17 & 6.981	\\      
Fe I & 605.6005  & -0.46   & FMW          & 9.18   & 6.795  & 7.43  & 6.517 &  8.31 & 6.692 &  8.08 & 6.654	\\      
Fe I & 609.6664  & -1.93   & FMW          & --     & --     & 6.05  & 6.707 &  6.19 & 6.748 &  9.76 & 7.324	\\      
Fe I & 615.1617  & -3.30   & FMW          & 18.60  & 7.279  &17.79  & 7.166 & 19.23 & 7.439 & 14.43 & 6.730	\\      
Fe I & 622.6734  & -2.22   & FMW          & --     & --     & 8.13  & 7.162 & --    & --    & --    & -- \\	        
Fe I & 651.8366  & -2.75   & FMW          & 11.27  & 6.580  &13.18  & 6.868 & 11.94 & 6.734 & 13.97 & 7.052	\\      
Fe I & 659.7559  & -1.07   & FMW          &  7.05  & 7.145  & 4.28  & 6.695 &  7.44 & 7.233 &  6.97 & 7.158	\\      
Fe I & 670.3566  & -3.16   & FMW          &  9.40  & 6.598  & 9.03  & 6.546 & 13.87 & 7.319 & --    & -- \\	        
Fe I & 673.9521  & -4.95   & FMW          & 11.68  & 6.885  &10.37  & 6.704 & 11.27 & 6.873 & 12.58 & 7.066	\\      
Fe I & 674.6954  & -4.35   & FMW	  & --     & --     & 4.74  & 6.923 &  6.99 & 7.254 &  5.55 & 7.052	\\      
Fe I & 679.3258  & -2.47   & FMW	  &  3.75  & 6.989  & 3.12  & 6.872 & --    & --    &  5.08 & 7.224	\\      
\\
Fe I & 595.6694  & -4.60   & FMW	  & --     & --     & 20.24 & 7.026 & 20.59 & 7.225 & --    & --	  \\	 	
Fe I & 595.8333  & -4.42   & K94	  &  9.36  & 7.040  & 10.36 & 7.187 & --    & --    & --    & --	     \\  	
Fe I & 602.4058  & -0.12   & FMW	  & 17.44  & 7.486  & 14.06 & 6.988 & 11.71 & 6.661 & 13.83 & 7.016 \\   	
Fe I & 606.5482  & -1.53   & FMW	  & 25.18  & 6.994  & 21.92 & 6.673 & 20.84 & 6.620 & 23.89 & 6.938 \\   	
Fe I & 624.6318  & -0.96   & FMW	  & 19.52  & 7.245  & 14.82 & 6.548 & 16.37 & 6.872 & 15.48 & 6.729 \\   	
Fe I & 625.2555  & -1.69   & FMW	  & --     & --     & 23.13 & 6.629 & 22.50 & 6.631 & 23.18 & 6.700 \\   	
Fe I & 629.7793  & -2.74   & FMW	  & --     & --     & 18.29 & 6.714 & 19.64 & 6.967 & 17.54 & 6.685 \\   	
Fe I & 630.1500  & -0.67   & K94	  & --     & --     & 14.88 & 6.335 & 15.81 & 6.560 & --    & --  \\	 	
Fe I & 630.2494  & -1.13   & K94	  & 12.39  & 6.442  & 13.55 & 6.629 & 17.73 & 7.360 & 16.04 & 7.106 \\   	
Fe I & 632.2685  & -2.43   & FMW	  & 15.38  & 6.551  & 14.90 & 6.476 & 17.76 & 7.001 & 17.16 & 6.909 \\   	
Fe I & 633.5330  & -2.23   & FMW	  & 25.32  & 6.923  & 22.29 & 6.643 & 22.59 & 6.733 & 21.80 & 6.654 \\   	
Fe I & 633.6823  & -1.05   & FMW	  & 19.20  & 7.397  & 16.53 & 7.016 & 16.27 & 7.054 & 15.30 & 6.900 \\   	
Fe I & 657.4227  & -5.04   & FMW	  & 18.29  & 7.107  & 17.54 & 6.990 & 17.25 & 7.037 & --    & --  \\	 	
Fe I & 659.3870  & -2.42   & FMW	  & 21.62  & 7.152  & 19.62 & 6.893 & 20.61 & 7.106 & 18.94 & 6.882 \\   	
Fe I & 664.8080  & -5.29   & K94	  & --     & --     & 11.18 & 6.325 & --    & --    & 12.35 & 6.538  \\  	
Fe I & 671.0318  & -4.88   & FMW	  & 11.55  & 6.686  & 13.05 & 6.900 & 10.81 & 6.627 & 12.96 & 6.946 \\   	
Fe I & 675.0152  & -2.62   & FMW	  & 20.26  & 7.233  & 17.74 & 6.829 & 18.33 & 7.024 & 18.84 & 7.109 \\   	
Fe I & 680.6843  & -3.21   & FMW	  & 10.46  & 6.744  & 11.22 & 6.853 & 13.74 & 7.287 & 13.32 & 7.222 \\   	
Fe I & 683.9830  & -3.45   & FMW	  & 12.10  & 6.963  & 11.26 & 6.842 & 14.43 & 7.374 & 12.47 & 7.071 \\   	
Fe I & 722.3658  & -2.21   & O  	  & 16.91  & 7.048  & --    &  --   & 18.22 & 7.291 & 13.20 & 6.584 \\   	
Fe I & 756.8899  & -0.87   & K94	  & 13.63  & 7.114  & 11.48 & 6.809 & 12.45 & 7.000 & 13.39 & 7.136 \\   	
Fe I & 758.3788  & -1.99   & FMW	  & 15.44  & 6.590  & 17.23 & 6.836 & 16.66 & 6.829 & 19.54 & 7.208 \\   	
Fe I & 774.8269  & -1.76   & FMW	  & 21.39  & 6.989  & 17.32 & 6.488 & 19.73 & 6.868 & 19.91 & 6.890 \\   	
Fe I & 783.2196  & -0.02   & K94	  & 15.37  & 6.719  & 14.80 & 6.641 & --    & --    & 16.70 & 6.962 \\   	
\\
Mg I & 552.8405  & -0.52   & G03          & 29.56  &  7.055 & 22.58 & 6.631 & 23.07 & 6.697 & 25.11 & 6.831\\   
Mg I & 571.1088  & -1.73   & G03          & 13.92  &  6.848 & 11.55 & 6.519 & 14.39 & 6.956 & 15.21 & 7.069\\   
Mg I & 631.8717  & -1.94   & G03          &  5.87  &  7.037 &  4.83 & 6.884 & --    & --    &  9.15 & 7.521\\   
Mg I & 631.9237  & -2.16   & G03          &  6.78  &  7.386 &  5.26 & 7.169 & --    & --    &  5.28 & 7.183\\   
\\
Ca I & 585.7451  &  0.24   & SR           & 18.52  &  5.373 & 17.00 & 5.171 & 19.85 & 5.596 & --    & --   \\   
Ca I & 586.7562  & -1.49   & G03          &  6.07  &  5.356 &  6.59 & 5.424 &  6.59 & 5.439 &  7.31 & 5.535\\   
Ca I & 643.9075  &  0.39   & SR           &    --  &	 -- & 22.20 & 5.111 & 24.50 & 5.474 & 25.39 & 5.565\\   
Ca I & 645.5558  & -1.29   & SR           & 14.51  &  5.654 & 12.81 & 5.403 & 13.25 & 5.523 & 12.34 & 5.385\\   
Ca I & 649.3781  & -0.11   & SR           & 21.04  &  5.423 & 21.85 & 5.533 & 23.21 & 5.795 & 22.20 & 5.669\\   
Ca I & 649.9650  & -0.82   & SR           & 14.09  &  5.110 & 16.81 & 5.512 &  --   & --    & 16.02 & 5.466\\   
Ca I & 650.8850  & -2.11   & NBS          &  6.57  &  5.402 &  4.86 & 5.180 &  6.27 & 5.377 &  5.66 & 5.297\\   
Ca I & 714.8150  &  0.21   & K88          & 21.33  &  5.288 & 24.59 & 5.694 & 21.31 & 5.367 & 24.13 & 5.721\\    
\\
\hline
\\
\\
\multispan{10}{FMW -- Fuhr et al.~(1988)\hfill}\\
\multispan{10}{G03 -- Gratton et al.~(2003)\hfill}\\
\multispan{10}{SR -- Smith et al.~(1981)\hfill}\\
\multispan{10}{NBS -- Wiese et al.~(1969)\hfill}\\
\multispan{10}{MFW -- Martin et al.~(1988)\hfill}\\
\multispan{10}{O   -- O'Brian et al.~(1991)\hfill}\\
\multispan{10}{K88   -- Kurucz~(1988)\hfill}\\
\multispan{10}{K94   -- Kurucz~(1994)\hfill}\\
\end{tabular}
}
\end{center}
\end{table*}

\addtocounter{table}{-1}

\begin{table*}
\begin{center}
\caption{Line list and adopted atomic parameters for the program stars. The measured equivalent width and the
corresponding abundance obtained for each line are also reported (continued).}
\label{abund1bis}
\scriptsize{
\begin{tabular}{rrrlrrrrrrrr}
\hline
\\
Ion  & $\lambda$ & log gf  & source of   & EW       &$\epsilon$ & EW       & $\epsilon$& EW (pm) & $\epsilon$ & EW      & $\epsilon$ \\
     & (nm)      &         & log gf      & (pm)     &           & (pm)     &           & (pm)    &            & (pm)    &            \\
     &           &         & (see notes) & \object{39}       &           & \object{42}       &           & \object{77}      &            & \object{232}&		     \\
\\
\hline
\\
Fe I & 585.5076  & -1.76   & FMW          & --    & --    & --    & --    &  3.42 & 6.991 & &	  \\
Fe I & 588.3817  & -1.36   & FMW          & 10.18 & 6.982 & 11.63 & 6.999 &  8.12 & 6.496 & --    & --    \\
Fe I & 595.2718  & -1.44   & FMW          &  8.66 & 6.795 & --    & --    &  6.50 & 6.337 &  9.61 & 6.774 \\
Fe I & 602.7051  & -1.21   & FMW          &  9.63 & 6.879 &  8.97 & 6.570 & 11.12 & 7.032 & 10.26 & 6.776 \\
Fe I & 605.6005  & -0.46   & FMW          &  9.97 & 7.127 &  8.48 & 6.684 &  8.85 & 6.822 & --    & --	  \\
Fe I & 609.6664  & -1.93   & FMW          & --    & --    &  6.51 & 6.777 & --    & --    & --    & --	  \\
Fe I & 615.1617  & -3.30   & FMW          & 16.41 & 7.282 & 14.11 & 6.610 & 12.70 & 6.517 & --    & --	  \\
Fe I & 622.6734  & -2.22   & FMW          &  6.88 & 7.085 &  7.55 & 7.075 &  7.43 & 7.115 & --    & --	  \\
Fe I & 651.8366  & -2.75   & FMW          & 13.11 & 7.126 & 13.62 & 6.934 & 13.14 & 6.987 & --    & --	  \\
Fe I & 659.7559  & -1.07   & FMW          &  8.21 & 7.473 &  5.27 & 6.862 &  4.66 & 6.790 &  8.62 & 7.387 \\
Fe I & 670.3566  & -3.16   & FMW          & 14.04 & 7.574 &  9.33 & 6.589 &  8.65 & 6.558 & 12.34 & 7.025 \\
Fe I & 673.9521  & -4.95   & FMW          & 10.13 & 6.849 &  9.69 & 6.614 &  6.12 & 6.188 & 10.51 & 6.723 \\
Fe I & 674.6954  & -4.35   & FMW	  &  4.59 & 6.946 &  2.75 & 6.588 & --    & --    &  7.66 & 7.324 \\
Fe I & 679.3258  & -2.47   & FMW	  &  4.52 & 7.175 & --    & --    &  2.28 & 6.699 &  3.49 & 6.942 \\
\\
Fe I & 595.6694  & -4.60   & FMW	  & --    & --    & --    & --    & --    & --    & 18.39 & 6.684 \\ 
Fe I & 595.8333  & -4.42   & K94	  & --    & --    &  9.58 & 7.072 &  8.90 & 7.045 & --    & --    \\ 
Fe I & 602.4058  & -0.12   & FMW	  & 12.20 & 6.947 & 14.65 & 7.081 & 11.99 & 6.772 & --    & --    \\ 
Fe I & 606.5482  & -1.53   & FMW	  & 23.51 & 7.047 & 24.52 & 6.935 & 18.41 & 6.380 & --    & --    \\ 
Fe I & 624.6318  & -0.96   & FMW	  & 16.04 & 7.065 & 13.81 & 6.383 & 12.91 & 6.372 & 18.17 & 7.060 \\ 
Fe I & 625.2555  & -1.69   & FMW	  & 24.06 & 6.941 & 20.18 & 6.272 & 20.73 & 6.503 & --    & --    \\ 
Fe I & 629.7793  & -2.74   & FMW	  & 18.36 & 7.022 & 16.60 & 6.474 & 17.70 & 6.784 & 21.40 & 7.103 \\ 
Fe I & 630.1500  & -0.67   & K94	  & 15.46 & 6.749 & --    & --    & --    & --    & 20.32 & 7.119 \\ 
Fe I & 630.2494  & -1.13   & K94	  & 11.82 & 6.609 & 14.14 & 6.724 & 11.32 & 6.384 & 15.62 & 6.961 \\
Fe I & 632.2685  & -2.43   & FMW	  & 16.06 & 6.995 & 15.42 & 6.557 & 14.43 & 6.553 & 20.26 & 7.268 \\
Fe I & 633.5330  & -2.23   & FMW	  & 22.17 & 6.862 & 20.51 & 6.443 & 21.28 & 6.664 & 25.64 & 6.949 \\
Fe I & 633.6823  & -1.05   & FMW	  & 12.76 & 6.699 & 15.87 & 6.914 & 14.54 & 6.853 & 15.85 & 6.911 \\
Fe I & 657.4227  & -5.04   & FMW	  & 17.28 & 7.375 & 15.65 & 6.694 & 15.35 & 6.809 & 17.16 & 6.928 \\
Fe I & 659.3870  & -2.42   & FMW	  & 18.75 & 7.107 & 18.63 & 6.751 & 17.99 & 6.830 & 20.47 & 7.006 \\ 
Fe I & 664.8080  & -5.29   & K94	  & --    & --    & --    & --    & --    & --    & --    & --    \\   
Fe I & 671.0318  & -4.88   & FMW	  & 12.46 & 7.067 & 11.02 & 6.614 &  9.63 & 6.499 & 13.63 & 6.985 \\ 
Fe I & 675.0152  & -2.62   & FMW	  & 17.57 & 7.223 & 17.65 & 6.814 & 15.25 & 6.596 & 17.93 & 6.860 \\ 
Fe I & 680.6843  & -3.21   & FMW	  & 12.14 & 7.230 &  8.98 & 6.536 & 11.60 & 7.012 & 15.19 & 7.440 \\ 
Fe I & 683.9830  & -3.45   & FMW	  & 10.49 & 6.927 & 10.57 & 6.743 & 10.53 & 6.826 & 12.84 & 7.070 \\ 
Fe I & 722.3658  & -2.21   & O  	  & 12.18 & 6.612 & 16.80 & 7.032 & 13.09 & 6.628 & --    & --  \\  
Fe I & 756.8899  & -0.87   & K94	  & 11.28 & 6.985 &  9.63 & 6.543 &  9.38 & 6.581 & 14.54 & 7.242 \\ 
Fe I & 758.3788  & -1.99   & FMW	  & 16.86 & 7.084 & 13.43 & 6.310 & 17.23 & 6.984 & 18.67 & 7.025 \\ 
Fe I & 774.8269  & -1.76   & FMW	  & 17.63 & 6.825 & 19.17 & 6.725 & --    & --    & 21.35 & 6.985 \\ 
Fe I & 783.2196  & -0.02   & K94	  & 15.51 & 7.000 & 14.60 & 6.613 & 11.90 & 6.334 & 15.55 & 6.744 \\ 
\\
Mg I & 552.8405  & -0.52   & G03          & 21.73 & 6.681 & --    & --    & 25.85 & 6.904 & 35.63 & 7.285\\
Mg I & 571.1088  & -1.73   & G03          & 11.84 & 6.714 & 13.58 & 6.801 & 14.69 & 7.043 & 21.27 & 7.702\\
Mg I & 631.8717  & -1.94   & G03          &  5.25 & 6.993 &  5.78 & 7.024 &  5.58 & 7.019 &  8.01 & 7.337\\
Mg I & 631.9237  & -2.16   & G03          & --    & --    &  3.91 & 6.957 &  4.32 & 7.040 &  9.10 & 7.709\\
\\
Ca I & 585.7451  &  0.24   & SR           & --    & --    & 18.84 & 5.412 & --    & --    & --    & --   \\
Ca I & 586.7562  & -1.49   & G03          &  6.30 & 5.452 &  6.57 & 5.421 &  6.05 & 5.380 &  9.29 & 5.778\\
Ca I & 643.9075  &  0.39   & SR           & 23.35 & 5.560 & 26.09 & 5.566 & 24.89 & 5.582 & 28.62 & 5.791\\
Ca I & 645.5558  & -1.29   & SR           & 11.66 & 5.445 & 10.51 & 5.084 & 11.26 & 5.273 & 14.07 & 5.588\\
Ca I & 649.3781  & -0.11   & SR           & 21.88 & 5.869 & 18.99 & 5.127 & 24.52 & 6.011 & 29.15 & 6.299\\
Ca I & 649.9650  & -0.82   & SR           & 15.63 & 5.644 & 18.92 & 5.829 & 16.29 & 5.587 & 19.82 & 5.961\\
Ca I & 650.8850  & -2.11   & NBS          &  5.31 & 5.286 &  4.92 & 5.188 & 10.40 & 5.961 & 12.66 & 6.198\\
Ca I & 714.8150  &  0.21   & K88          & 21.63 & 5.668 & 24.18 & 5.646 & --    & --    & 27.45 & 5.991\\ 
\\
\hline
\\
\\
\multispan{10}{FMW -- Fuhr et al.~(1988)\hfill}\\
\multispan{10}{G03 -- Gratton et al.~(2003)\hfill}\\
\multispan{10}{SR -- Smith et al.~(1981)\hfill}\\
\multispan{10}{NBS -- Wiese et al.~(1969)\hfill}\\
\multispan{10}{MFW -- Martin et al.~(1988)\hfill}\\
\multispan{10}{O   -- O'Brian et al.~(1991)\hfill}\\
\multispan{10}{K88   -- Kurucz~(1988)\hfill}\\
\multispan{10}{K94   -- Kurucz~(1994)\hfill}\\
\end{tabular}
}
\end{center}
\end{table*}

\addtocounter{table}{-1}

\begin{table*}
\begin{center}
\caption{Line list and adopted atomic parameters for the program stars. The measured equivalent width and the
corresponding abundance obtained for each line are also reported (continued).}
\label{abund1ter}
\scriptsize{
\begin{tabular}{rrrlrrrrrrrr}
\hline
\\
Ion  & $\lambda$ & log gf  & source of   & EW       &$\epsilon$ & EW       & $\epsilon$& EW (pm) & $\epsilon$ & EW      & $\epsilon$ \\
     & (nm)      &         & log gf      & (pm)     &           & (pm)     &           & (pm)    &            & (pm)    &            \\
     &           &         & (see notes) & \object{242}      &           & \object{278}      &           & \object{281}     &            & \object{283}&		     \\
\\
\hline
\\
Fe I & 585.5076  & -1.76   & FMW          &  5.72 & 7.352 & --    & --    & --    & --    & --    & --   \\
Fe I & 588.3817  & -1.36   & FMW          & 11.53 & 6.927 &  9.43 & 6.677 & --    & --    & --    & --   \\
Fe I & 595.2718  & -1.44   & FMW          & --    & --    & --    & --    &  7.76 & 6.455 &  7.05 & 6.400\\
Fe I & 602.7051  & -1.21   & FMW          & 10.49 & 6.768 & --    & --    & --    & --    & 10.04 & 6.787\\
Fe I & 605.6005  & -0.46   & FMW          & 10.37 & 6.943 & --    & --    & --    & --    &  9.90 & 6.956\\
Fe I & 609.6664  & -1.93   & FMW          &  7.12 & 6.846 & --    & --    &  4.76 & 6.494 & --    & --   \\
Fe I & 615.1617  & -3.30   & FMW          & --    & --    & --    & --    & --    & --    & --    & --	 \\
Fe I & 622.6734  & -2.22   & FMW          & --    & --    & --    & --    & --    & --    & --    & --	 \\
Fe I & 651.8366  & -2.75   & FMW          & 14.65 & 7.025 & 12.76 & 6.863 &  9.72 & 6.320 & --    & --   \\
Fe I & 659.7559  & -1.07   & FMW          &  5.24 & 6.842 & --    & -- 	  & --    & --    & --    & --   \\
Fe I & 670.3566  & -3.16   & FMW          & 12.66 & 7.019 & 11.00 & 6.873 & 10.05 & 6.655 & 11.96 & 7.021\\
Fe I & 673.9521  & -4.95   & FMW          & 10.45 & 6.679 &  9.00 & 6.553 &  7.72 & 6.342 & 13.07 & 7.141\\
Fe I & 674.6954  & -4.35   & FMW	  & --    & --    & --    & --    &  5.22 & 6.981 & --    & --   \\
Fe I & 679.3258  & -2.47   & FMW	  &  4.91 & 7.170 & --    & --    & --    & --    &  3.55 & 6.962\\
\\
Fe I & 595.6694  & -4.60   & FMW	  & 16.41 & 6.242 & --    & --    & 19.00 & 6.683 & 19.00 & 6.919\\ 
Fe I & 595.8333  & -4.42   & K94	  &  7.40 & 6.747 & --    & --    &  8.38 & 6.876 &  6.24 & 6.627\\ 
Fe I & 602.4058  & -0.12   & FMW	  & 12.83 & 6.728 & --    & --    & 12.05 & 6.603 & 10.68 & 6.486\\ 
Fe I & 606.5482  & -1.53   & FMW	  & --    & --    & 23.45 & 6.895 & 21.45 & 6.542 & 21.61 & 6.708\\ 
Fe I & 624.6318  & -0.96   & FMW	  & 15.55 & 6.589 & --    & --    & 15.87 & 6.639 & --    & --   \\ 
Fe I & 625.2555  & -1.69   & FMW	  & --    & --    & 26.25 & 6.971 & --    & --    & 24.47 & 6.823\\ 
Fe I & 629.7793  & -2.74   & FMW	  & 15.56 & 6.248 & 21.39 & 7.168 & --    & --    & 15.42 & 6.369\\ 
Fe I & 630.1500  & -0.67   & K94	  & 16.14 & 6.459 & 15.86 & 6.568 & 17.10 & 6.605 & 15.84 & 6.565\\ 
Fe I & 630.2494  & -1.13   & K94	  & 13.02 & 6.483 & 12.69 & 6.553 & 12.64 & 6.425 & 13.47 & 6.684\\
Fe I & 632.2685  & -2.43   & FMW	  & 15.12 & 6.441 & 15.88 & 6.708 & 15.14 & 6.444 & 15.07 & 6.578\\
Fe I & 633.5330  & -2.23   & FMW	  & --    & --    & 24.91 & 6.940 & 20.67 & 6.391 & 22.60 & 6.734\\
Fe I & 633.6823  & -1.05   & FMW	  & 13.71 & 6.505 & --    & --    & 13.92 & 6.540 & 12.88 & 6.501\\
Fe I & 657.4227  & -5.04   & FMW	  & 14.60 & 6.471 & 18.63 & 7.268 & 16.10 & 6.687 & 15.88 & 6.810\\
Fe I & 659.3870  & -2.42   & FMW	  & 17.24 & 6.469 & 19.73 & 6.990 & 17.47 & 6.503 & 17.39 & 6.655\\ 
Fe I & 664.8080  & -5.29   & K94	  & --    & --    & 11.28 & 6.383 & --    & --    & 12.07 & 6.497\\ 	  
Fe I & 671.0318  & -4.88   & FMW	  &  9.44 & 6.374 &  9.86 & 6.493 &  8.40 & 6.246 & 11.99 & 6.799\\ 
Fe I & 675.0152  & -2.62   & FMW	  & 14.22 & 6.206 & 16.74 & 6.758 & 16.87 & 6.605 & 15.44 & 6.541\\ 
Fe I & 680.6843  & -3.21   & FMW	  & 11.65 & 6.869 & --    & --    & 10.22 & 6.673 & 11.97 & 7.014\\ 
Fe I & 683.9830  & -3.45   & FMW	  & --    & --    & --    & --    & --    & --    & --    & --   \\ 
Fe I & 722.3658  & -2.21   & O  	  & --    & --    & --    & --    & 11.75 & 6.278 & 12.76 & 6.520\\ 
Fe I & 756.8899  & -0.87   & K94	  & 10.44 & 6.620 & --    & --    &  8.56 & 6.361 & 10.48 & 6.707\\ 
Fe I & 758.3788  & -1.99   & FMW	  & 16.06 & 6.611 & --    & --    & 14.70 & 6.429 & --    & --   \\ 
Fe I & 774.8269  & -1.76   & FMW	  & 16.95 & 6.370 & 20.60 & 6.971 & 19.16 & 6.652 & --    & --   \\ 
Fe I & 783.2196  & -0.02   & K94	  & --    & --    & --    &  --   & --    & --    & 15.40 & 6.786\\ 
\\
Mg I & 552.8405  & -0.52   & G03     	  & 27.28 & 6.913 & 23.13 & 6.701 & 24.86 & 6.765 & 23.99 & 6.759\\
Mg I & 571.1088  & -1.73   & G03     	  & 16.77 & 7.177 & --    & --    & 15.06 & 6.958 & 13.41 & 6.819\\
Mg I & 631.8717  & -1.94   & G03     	  &  6.01 & 7.045 &  5.10 & 6.935 &  2.93 & 6.554 &  6.12 & 7.086\\
Mg I & 631.9237  & -2.16   & G03     	  & --    & --    &  4.42 & 7.049 &  4.60 & 7.060 &  4.61 & 7.079\\
\\
Ca I & 585.7451  &  0.24   & SR 	  & --    & --    & --    & --    & --    & --    & --    & --   \\
Ca I & 586.7562  & -1.49   & G03     	  &  8.14 & 5.603 &  6.08 & 5.370 &  8.39 & 5.658 &  9.15 & 5.788\\
Ca I & 612.2217  & -0.31   & NIST      	  & --    & --    & --    & --    & --    & --    & 26.64 & 5.527\\
Ca I & 643.9075  &  0.39   & SR 	  & 25.02 & 5.378 & 24.47 & 5.471 & 27.48 & 5.693 & 22.89 & 5.286\\
Ca I & 645.5558  & -1.29   & SR 	  & 13.46 & 5.445 & 11.82 & 5.310 & 15.41 & 5.788 & 10.51 & 5.120\\
Ca I & 649.3781  & -0.11   & SR 	  & 21.55 & 5.408 & 22.55 & 5.713 & 21.13 & 5.436 & --    & --   \\
Ca I & 649.9650  & -0.82   & SR 	  & 15.81 & 5.297 & 17.32 & 5.667 & 14.56 & 5.179 & --    & --   \\
Ca I & 650.8850  & -2.11   & NBS     	  &  7.85 & 5.542 &  7.30 & 5.510 &  9.29 & 5.742 &  5.03 & 5.212\\
Ca I & 714.8150  &  0.21   & K88     	  & 25.94 & 5.763 & 24.47 & 5.761 & 22.89 & 5.486 & 22.59 & 5.535\\ 
\\
\hline
\\
\multispan{10}{FMW -- Fuhr et al.~(1988)\hfill}\\
\multispan{10}{G03 -- Gratton et al.~(2003)\hfill}\\
\multispan{10}{SR -- Smith et al.~(1981)\hfill}\\
\multispan{10}{NIST -- Furh \& Wiese~(1996)\hfill}\\
\multispan{10}{NBS -- Wiese et al.~(1969)\hfill}\\
\multispan{10}{MFW -- Martin et al.~(1988)\hfill}\\
\multispan{10}{O   -- O'Brian et al.~(1991)\hfill}\\
\multispan{10}{K88   -- Kurucz~(1988)\hfill}\\
\multispan{10}{K94   -- Kurucz~(1994)\hfill}\\
\end{tabular}
}
\end{center}
\end{table*}

\end{document}